\documentclass[review]{elsarticle}

\usepackage{lineno,hyperref}
\usepackage{amsmath, amssymb, amsthm}
\usepackage{soul}
\usepackage{mathtools}
\usepackage{graphics}
\usepackage{graphicx}
\usepackage{epsfig}
\usepackage{dblfloatfix}
\usepackage{float}

\newtheorem{theorem}{\bf Theorem}
\newtheorem{lemma}{\bf Lemma}

\newtheorem{example}{\bf Example}
\newtheorem{remark}{\bf Remark}

\usepackage{epstopdf}
\usepackage{url}
\usepackage{xcolor}
\usepackage{hyperref}
\date{}
\usepackage{fullpage}
\usepackage{setspace}
\usepackage{lscape}

\newtheorem{corollary}{Corollary}[section]

\journal{Submitted to Discrete Mathematics}

\begin{document}
	
	\begin{frontmatter}
		
		\title{Galois hulls of constacyclic codes over affine algebra rings}
		
		
		\author{Indibar Debnath}
		\address{Department of Mathematics, Indian Institute of Technology Patna, India.}
		\ead{indibar\_1921ma07@iitp.ac.in}

        \author{Habibul Islam}
		\address{Department of Mathematics, Indian Institute of Information Technology Bhopal, India.}		\ead{habibul.islam@iiitbhopal.ac.in}
        
		\author{E. Mart\'inez-Moro}
		\address{Institute of Mathematics, University of Valladolid, Castilla, Spain.}
		\ead{edgar.martinez@uva.es}
		
		\author{Om Prakash\corref{mycorrespondingauthor}}
		\address{Department of Mathematics, Indian Institute of Technology Patna, India.}
		\cortext[mycorrespondingauthor]{Corresponding author}
		\ead{om@iitp.ac.in}

		\begin{abstract}
		Let $\mathcal A$ the affine algebra  given by the ring $\mathbb{F}_q[X_1,X_2,\ldots,X_\ell]/ I$, where $I$ is the ideal $\langle t_1(X_1), t_2(X_2), \ldots, t_\ell(X_\ell) \rangle$ with each $t_i(X_i)$, $1\leq i\leq \ell$, being a square-free polynomial over $\mathbb{F}_q$. This paper studies the $k$-Galois hulls of $\lambda$-constacyclic codes over $\mathcal A$ regarding their idempotent generators. For this, first, we define the $k$-Galois inner product over $\mathcal A$ and find the form of the generators of the $k$-Galois dual and the $k$-Galois hull of a $\lambda$-constacyclic code over $\mathcal A$. Then, we derive a formula for the $k$-Galois hull dimension of a $\lambda$-constacyclic code. Further, we provide a condition for a $\lambda$-constacyclic code to be $k$-Galois LCD. Finally, we give some examples of the use of these codes in constructing entanglement-assisted quantum error-correcting codes.
		\end{abstract}
		
		
	\end{frontmatter}
	
	
	\section{Introduction}
	The intersection of a linear code with its dual is known as the hull of that code, and it was introduced in \cite{Assmus} related to some properties of finite projective planes. Over the years, researchers have discovered various applications of the hull. The dimension of the hull has an important role in examining the complexities of the algorithms that are used to check the permutation equivalence of two linear codes and determining the automorphism group of a linear code since these algorithms fail when the dimension of the hull is large \cite{Leon1,Leon2,Leon3,Petrank,Sendrier2}. In addition, the hull of linear codes is used to obtain entanglement-assisted quantum error-correcting codes \cite{Dougherty,Tian}. Note that a linear code with a zero-dimensional hull is known as a linear complementary dual (LCD) code. LCD codes have been used in secret sharing schemes \cite{Yadav}, and in cryptosystems as protection against side-channel attacks and fault injection attacks \cite{Carlet1}. 

Sendrier \cite{Sendrier4} studied the hull dimensions of linear codes and counted the number of $q$-ary linear codes of a given hull dimension. Later, in $2003$, Skersys \cite{Skersys} studied the average dimension of the hull for the very first time. He derived an expression for the average hull dimensions of cyclic codes over finite fields. Sangwisut et al. \cite{Sangwisut2}, in $2015$, proposed formulas for the dimensions of the hulls for both cyclic and negacyclic codes and calculated the number of codes for a given dimension of the hull. Jitman et al. \cite{Jitman} considered constacyclic codes and derived a formula for the average dimension of their Hermitian hull over finite fields of square order.

In 2017, the Galois inner product was introduced over finite fields in \cite{Fan}. This product generalizes both the Euclidean and the Hermitian inner products. In \cite{LiuH}, the Galois hulls of linear codes in finite fields were studied. Meanwhile, Ding et al. \cite{Ding} studied the Galois hulls of cyclic codes over finite fields. Cao \cite{Cao} and Li et al. \cite{LiY} constructed MDS codes from arbitrary dimensions of the Galois hull. In $2023$, Debnath et al. \cite{Indibar} presented a formula to calculate the Galois hull dimensions of constacyclic codes over finite fields. They also counted the number of constacyclic codes of a given Galois hull dimension.

The study of codes over finite rings started in the early 1970s, and many optimal codes are obtained by considering different finite rings as the code alphabets. In \cite{Martinez-Moro} it was established the structure of multivariate serial codes over a finite chain ring, and the authors investigated the structure of the dual codes in the serial Abelian case. In $2020$, Goyal and Raka  \cite{Goyal}  studied polyadic constacyclic codes over the commutative ring $\mathbb{F}_q[u,v]/\langle f(u), g(v)\rangle$, where   $f(u)$ and $g(v)$are two polynomials that split into linear factors over $\mathbb{F}_q$. In $2021$, Islam et al. \cite{Islam1} studied the cyclic codes over a non-chain ring $R_{e,q}$. They gave some necessary and sufficient conditions on cyclic codes over $R_{e,q}$ to be LCD and obtained many MDS cyclic codes as well as optimal LCD codes. Prakash et al. \cite{Prakash} presented the study of self-dual and LCD double circulant codes over a non-chain ring. In $2024$, Y\i lmazg\"{u}\c{c} et al. \cite{Yilmaz} described the structure of the Abelian, consta-Abelian and polyadic codes over a polynomial serial ring.

Although hulls of different families of linear codes have been studied over finite fields, only a few works are available in the literature on hulls over finite rings. Jitman and Sangwisut \cite{Jitman3} determined the hull dimension and the average hull dimension of cyclic codes over the finite non-chain ring $\mathbb{F}_2[v]/\langle v^2-v\rangle$. Further, Jitman et al. \cite{Jitman2} studied the hulls of cyclic codes over $\mathbb{Z}_4$ in $2020$. Talbi et al. \cite{Talbi}, in $2022$, explored the hulls of cyclic serial codes over a finite chain ring and established the average dimensions of the Euclidean hulls of those codes. In $2023$, Dougherty and Salt\"{u}rk \cite{Dougherty2} investigated the hulls of both additive and linear codes over some finite rings of order $4$ and enumerated these codes having hull of a given type. In $2024$, Yadav et al. \cite{Yadav2} studied the Hermitian hulls of constacyclic codes over a more general class of non-chain ring $\mathbb{F}_q[u]/\langle  u^e-1\rangle$ such that $\mathrm{gcd}(e,q)=1$. Besides finding the Hermitian hull dimensions of $\lambda$-constacyclic codes over the ring, they also obtained the conditions for these codes to be Hermitian LCD.

Motivated by the above works, in this paper given $\mathbb{F}_q$ a finite field with $q=p^e$ number of elements, where $p$ is a prime and $e$ is a positive integer, and  $t_i(X_i)$  a square-free polynomial in $\mathbb{F}_q[X_i]$ for $i=1,2,\ldots,\ell$, we will consider the  commutative affine algebra ring 
\begin{equation}\label{eq:algebraA}
\mathcal A = \mathbb{F}_q[X_1,X_2,\ldots,X_{\ell}]/ I    \hbox{ where } I=\langle t_1(X_1), t_2(X_2), \ldots, t_{\ell}(X_{\ell}) \rangle
\end{equation}
 as the code alphabet. Note that this type of algebras comprises, as particular examples, several of the rings that already were studied related to the construction of constacyclic codes, such as the two mentioned above $\mathbb{F}_2[v]/\langle v^2-v\rangle$ \cite{Jitman3}, $\mathbb{F}_q[u]/\langle  u^e-1\rangle$ when $\mathrm{gcd}(e,q)=1$ \cite{Yadav2,Alahmadi}, and also $\mathbb F_q [u, v]/\langle u^2-1, v^2 - v\rangle$ in \cite{Gao}, $R_{k,m} = \mathbb F_{p^m} [u_1, u_2, \ldots , u_k ]/\langle u^2_i -1\rangle_{1\leq i\leq k}$ in \cite{Habibul}. More generally, when dealing with polycyclic codes, this type of ambient alphabet has been studied from a transform approach point of view in \cite{Bajalan_2022,Bajalan_2025}. 
 
 Our main objective is to derive a formula for the $q$-dimensions of the $k$-Galois hulls of $\lambda$-constacyclic codes over a serial ring $\mathcal A$. For this purpose, first, we decompose the ring $\mathcal A$ as a direct sum of simple ideals. Using this decomposition, we find the form of generators of constacyclic codes, their Galois duals, and Galois hulls. Then, with the help of the factorization of $x^n-\lambda$ over $\mathbb{F}_q$ in which the irreducible factors are arranged in a suitable way as given in \cite{Indibar}, we derive an expression for the Galois hull dimensions of constacyclic codes over $\mathcal A$. Our study will rely on the factorization of the defining constacyclic polynomial (as a $\mathbb F_q$ polynomial)  when projected in each simple component copy of the decomposition of the ambient space. This will provide some properties that allow to compute the cardinality of the set of $q$-dimensions of $k$-Galois hulls of $\lambda$-constacyclic codes of length $n$ over $\mathcal A$, denoted as $\delta_k(n,q,\lambda)$. 
 
It is well-known that constructing entanglement-assisted quantum error-correcting codes (shortly, EAQECCs) does not require the orthogonality property of classical codes, unlike standard stabilizer codes \cite{Brun06}. However, the number of preshared entanglements between the sender and the receiver plays an important role in the construction of EAQECCs. Guenda et al. \cite{Guenda18} proved that this number can be derived from the hull dimension of classical codes. As a result, studying the hull dimension of several important classical codes becomes relevant in the context of EAQECCs. For example, Fang et al. \cite{Fang20} studied Euclidean and Hermitian hull of generalized Reed-Solomon codes and explored many new MDS EAQECCs. Recently, the Galois hull dimension of linear codes has been used by Cao \cite{Cao} in the context of EAQECCs. In fact, he provided nine families of new EAQECCs. In this article, we also use the Galois hull dimension of constacyclic codes over $\mathcal{A}$ and provide some new classes of EAQECCs. 

 
This paper is organized as follows. In Section~\ref{sec:pre} contains some preliminaries. We decompose the ring $\mathcal A$ as a direct sum of simple ideals and show the existence of the primitive orthogonal idempotent elements in $\mathcal A$. Further, we recall a few definitions and results on the Galois inner product, constacyclic codes over $\mathcal A$, and the factorization of $x^n-\lambda$ over $\mathbb{F}_q$. In Section~\ref{sec.3}, first, we find the form of generators of the Galois duals and the Galois hulls of constacyclic codes over $\mathcal A$. We present a formula for the dimensions of the Galois hulls of constacyclic codes over $\mathcal A$. Further, some conditions are provided on the constacyclic codes over $\mathcal A$ to be Galois self-dual, Galois dual-containing, and Galois LCD. Section~\ref{sec:quan} shows an application of our setting to the construction of entanglement-assisted quantum error-correcting codes over the field $\mathbb F_q$. We get some examples of MDS and Near MDS codes as well as Optimal and Near Optimal codes with respect to Grassl's Table~\cite{Grassl}.

\section{Preliminaries}\label{sec:pre}

Let $\mathbb{F}$ be a field. A linear code of length $n$ is an $\mathbb{F}$-subspace of $\mathbb{F}^n$. Let $\lambda$ be a unit in $\mathbb F^*$, a linear code $C\subseteq \mathbb F^n$  is said to be a $\lambda$-constacyclic code if $(c_0,c_1,\ldots,c_{n-1})\in C$ implies that $(\lambda c_{n-1},c_0,c_1,\ldots,c_{n-2})\in C$. It is well known that we can identify  a vector $\mathbf c\in \mathbb F^n$ with a polynomial $c_0+c_1x+\cdots+c_{n-1}x^{n-1}$ in $\frac{\mathbb F[x]}{\langle x^n-\lambda\rangle}$ and    $\lambda$-constacyclic codes are  identified with ideals in $\frac{\mathbb F[x]}{\langle x^n-\lambda\rangle}$. Abusing the notation, we will use both the polynomial notation or the vector space one for constacyclic codes during the paper. 

As usual,  the Hamming weight $w_H(x)$ of $x=(x_1,x_2,\dots,x_{n})\in  \mathbb{F}^n$ is $w_H(x)= ~\mid\{ i \mid x_i \neq 0 \}\mid$  and the Hamming distance $d_H(x,y)$ between  $x=(x_1,x_2,\dots,x_{n}),\, y=(y_1,y_2,\dots,y_{n})\in  \mathbb{F}^n$ is  $d_H(x,y)= w_H(x-y)$. The minimum Hamming distance $d_H(C)$ of a linear code $C$ is defined by $d_H(C)=min\{d_H(x,y) \mid x,y \in C,~ x \neq y \}.$

We use the notation $[n,\mathrm{k},d]_q$ to represent a linear code of length $n$, dimension $\mathrm{k}$, and minimum Hamming distance $d$ over a finite field $\mathbb{F}_q$.

\subsection{The code alphabet and its decomposition}\label{Section2.1}

Let $\mathcal A$ be an affine algebra defined as in Equation~(\ref{eq:algebraA}). It is well known, see \cite{Poli} or more generally  \cite{Martinez-Moro} for affine algebras over chain rings and the proofs of the statements in this subsection, that $\mathcal A$ is a ring with a decomposition given by primitive idempotents related to the roots of the generators of the ideal $I$. More precisely, for each $i=1,2,\ldots,\ell$, we denote the set of roots of $t_i(X_i)$ in an extension field of $\mathbb F_q$ where the polynomial $t_i(X_i)$ splits into linear factors by $H_i$, and let $H = \prod_{i=1}^{\ell} H_i$. Then for each $\nu = (\nu_1,\nu_2,\ldots,\nu_{\ell})\in H$, one can define the class of $\nu$ as $S(\nu) = \{(\nu_1^{q^j},\nu_2^{q^j},\ldots,\nu_{\ell} ^{q^j}) \in H : j\in \mathbb{N}\}$. In this paper, the set of all those classes will be denoted by $\Hat{S}$ and its size by $\mathcal{N}$, i.e., $\mid\Hat{S}\mid~ = \mathcal{N}$. The elements of $\Hat{S}$ form a partition of $H$, and the size of each class is given by
$
    \mid S(\nu) \mid~ = \mathrm{lcm}(d_1,d_2,\ldots,d_{\ell}) = [\mathbb{F}_q(\nu_1,\nu_2,\ldots,\nu_{\ell}) : \mathbb{F}_q]
$, 
where $d_i$ is the degree of $\mathrm{Irr}(\nu_i,\mathbb{F}_q)$, the irreducible polynomial of $\nu_i$ over $\mathbb{F}_q$.

  We denote the polynomial $\mathrm{Irr}(\nu_i,\mathbb{F}_q)$ by $p_{S(\nu),i}(X_i)$ and, for each $i=2,3,\ldots,\ell$, we denote by $\mathrm{Irr}(\nu_i,\mathbb{F}_q(\nu_1,\nu_2,\ldots,\nu_{i-1}))$, the irreducible polynomial of $\nu_i$ over $\mathbb{F}_q(\nu_1,\nu_2,\ldots,\nu_{i-1})$ by $b_{S(\nu),i}(X_i)$. Applying the division algorithm on $p_{S(\nu),i}(X_i)$ and $b_{S(\nu),i}(X_i)$, we get that $b_{S(\nu),i}(X_i)$ divides $p_{S(\nu),i}(X_i)$. Note that the above polynomials do not depend on the representative of the class $S(\nu)$ chosen for their computation.
  Let 
$$\Tilde{b}_{S(\nu),i}(X_i) = \frac{p_{S(\nu),i}(X_i)}{b_{S(\nu),i}(X_i)}, \quad w_{S(\nu),i}(X_1,X_2,\ldots,X_i) \hbox{ and } \Tilde{w}_{S(\nu),i}(X_1,X_2,\ldots,X_i)\hbox{ for }  i=2,3,\ldots,\ell$$
be the multivariable polynomials obtained from $b_{S(\nu),i}(X_i)$ and $\Tilde{b}_{S(\nu),i}(X_i)$, respectively, by substituting $\nu_j$ by $X_j$; $j=1,2,\ldots,i-1$. Given the ideal $I_{S(\nu)}=\langle p_{S(\nu),1}, w_{S(\nu),2}, w_{S(\nu),3}, \ldots, w_{S(\nu),\ell} \rangle$, we have $\mathbb{F}_q[X_1,X_2,\ldots,X_{\ell}]/I_{S(\nu)} \cong \mathbb{F}_q(\nu_1,\nu_2,\ldots,\nu_{\ell})$, and by using the Chinese Remainder Theorem, we get
\begin{align*}
    R \cong \mathbb{F}_q[X_1,X_2,\ldots,X_{\ell}]/\bigcap_{S(\nu)\in \Hat{S}} I_{S(\nu)} \cong \bigoplus_{S(\nu)\in \Hat{S}} \mathbb{F}_q[X_1,X_2,\ldots,X_{\ell}]/ I_{S(\nu)} \cong \bigoplus_{S(\nu)\in \Hat{S}} \mathbb{F}_q(\nu_1,\nu_2,\ldots,\nu_{\ell}).
\end{align*}
Let us consider the polynomials
\begin{align*}
    h_{S(\nu)}(X_1,X_2,\ldots,X_{\ell}) = \prod_{i=1}^{\ell} \frac{t_i(X_i)}{p_{S(\nu),i}(X_i)} \prod_{i=2}^{\ell} \Tilde{w}_{S(\nu),i},
\end{align*}
 we have $I_{S(\nu)} + I = \mathrm{Ann}(\langle h_{S(\nu)} + I \rangle)$, $\langle h_{S(\nu)} + I \rangle \cong \mathbb{F}_q[X_1,X_2,\ldots,X_{\ell}]/ I_{S(\nu)}$, and $$\mathcal A \cong \bigoplus_{S(\nu)\in \Hat{S}} \langle h_{S(\nu)} + I \rangle.$$
This decomposition of the ring $\mathcal A$ is equivalent to the existence of a \emph{ set of primitive orthogonal idempotent elements} in $\mathcal A$ denoted by  $e_{S(\nu)}$ such that $\sum_{S(\nu)\in \Hat{S}} e_{S(\nu)} = 1$ and $e_{S(\nu)}R \cong \mathbb{F}_q(\nu_1,\nu_2,\ldots,\nu_{\ell})$. Moreover, each idempotent $e_{S(\nu)}$ is the element $g_{S(\nu)} h_{S(\nu)} + I$ such that $g_{S(\nu)} h_{S(\nu)} + I_{S(\nu)} = 1 + I_{S(\nu)}$. In other words,  this is equivalent to having a decomposition as a direct sum of finite fields, and thus, $\mathcal A$ is a finite commutative semi-simple ring. 

\begin{example}
    Let us consider the ring $\mathcal A_1 = \mathbb{F}_2[x_1,x_2]/\langle t_1(x_1), t_2(x_2) \rangle$, where $t_1(x_1) = x_1^5+x_1^4+1=(x_1^2+x_1+1)(x_1^3+x_1+1)$, $t_2(x_2) = x_2^4+x_2^2+x_2=x_2(x_2^2+x_2+1)$. It is easy to compute that, over a suitable field extension of $\mathbb F_2$, there are six equivalence conjugacy classes in the set of roots of $t_1(x_1), t_2(x_2)$. After dealing with the computations above, one can check that the primitive idempotents in $\mathcal A_1$ are given by  
    \begin{align*}
        e_{S_1}(x_1,x_2) =&~ x_1^4 x_2^3 + x_1^3 x_2^3 + x_1^4 x_2 + x_1^2 x_2^3 + x_1^4 + x_1^3 x_2 + x_1^3 + x_1^2 x_2 + x_2^3 + x_1^2 + x_2 + 1,\\
        e_{S_2}(x_1,x_2) =&~ x_1^4 x_2^3 + x_1^3 x_2^3 + x_1^4 x_2 + x_1^2 x_2^3 + x_1^3 x_2 + x_1^2 x_2 + x_2^3 + x_2,\\
        e_{S_3}(x_1,x_2) =&~ x_1^4 x_2^3 + x_1^3 x_2^3 + x_1^4 x_2 + x_1^2 x_2^3 + x_1^4 + x_1^3 x_2 + x_1^3 + x_1^2 x_2 + x_1^2,\\
        e_{S_4}(x_1,x_2) =&~ x_1^4 x_2^3 + x_1^3 x_2^3 + x_1^3 x_2^2 + x_1^2 x_2^3 + x_1^3 x_2 + x_1^2 x_2 + x_1x_2 + x_2^2,\\
        e_{S_5}(x_1,x_2) =&~ x_1^4 x_2^3 + x_1^4 x_2^2 + x_1^3 x_2^3 + x_1^2 x_2^3 + x_1^2 x_2 + x_1 x_2^2 + x_1 x_2 + x_2,\\
        e_{S_6}(x_1,x_2) =&~ x_1^4 x_2^3 + x_1^4 x_2^2 + x_1^3 x_2^3 + x_1^4 x_2 + x_1^3 x_2^2 + x_1^2 x_2^3 + x_1^2 x_2 + x_1 x_2^2 + x_2^2 + x_2.
    \end{align*}
\end{example}

\subsection{Galois inner product}\label{Section2.2}

The Galois inner product over $\mathbb{F}_q^n$ was introduced in \cite{Fan}. For any two elements $\boldsymbol{\alpha} = (\alpha_0,\alpha_1,\dots,\alpha_{n-1})$, $\boldsymbol{\beta} = (\beta_0,\beta_1,\dots,\beta_{n-1})$ in $\mathbb{F}_q^n$, and for any integer $k$, $0\leq k<e$ where $p^e=q$, the \emph{ $k$-Galois inner product} of $\boldsymbol{\alpha}$ and $\boldsymbol{\beta}$ is denoted by $\langle \boldsymbol{\alpha},\boldsymbol{\beta}\rangle_k$ is defined as
\begin{equation}
    \langle \boldsymbol{\alpha},\boldsymbol{\beta}\rangle_k =\sum_{i=0}^{n-1} \alpha_i\beta_i^{p^{k}}.
\end{equation}
Note that the Euclidean and Hermitian inner products are particular cases of the Galois inner product when $k=0$, and $k=\frac{e}{2}$ where $e$ is even, respectively. If $C\subseteq \mathbb{F}_q^n$ is a linear code, its \emph{$k$-Galois dual} is given by
$
    C^{\perp_k} =  \left\{\boldsymbol{\alpha} \in\mathbb{F}_q^n \mid \langle\boldsymbol{\alpha},\boldsymbol{c}\rangle_k = 0 ~\text{for all } \boldsymbol{c}\in C\right\}.
$
Note that the $k$-Galois dual $C^{\perp_k}$ is also a linear code of length $n$ and $\dim_{\mathbb{F}_q}(C) + \dim_{\mathbb{F}_q}(C^{\perp_k}) = n$ \cite{LiuX}. 

For a linear code $C$ the \emph{$k$-Galois hull} of $C$ is  the linear code $\mathrm{hull}_k(C) = C\cap C^{\perp_k}$.   In particular, $C$ is said to be a \emph{$k$-Galois linear complementary dual (LCD) code} if $\mathrm{hull}_k(C)=\{\boldsymbol{0}\}$, \emph{$k$-Galois dual containing} if $\mathrm{hull}_k(C)=C^{\perp_k}$, and  \emph{$k$-Galois self-orthogonal} if $\mathrm{hull}_k(C)=C$. 

 For a constacyclic code of length $n$, $C^{\perp_k}$ and $\mathrm{hull}_k(C)$ both become constacyclic codes under certain conditions, as shown in the following two results.
\begin{lemma}\cite{LiuX}\label{Lemma1}
    Let $\lambda \in \mathbb{F}_q^{\ast}$ and $\mathrm{ord}(\lambda) = r$. Then, we have the following.
\begin{enumerate}
    \item If $C$ is a $\lambda$-constacyclic code over $\mathbb{F}_q$, then $C^{\perp_k}$ is a $\lambda^{-p^{e-k}}$-constacyclic code over $\mathbb{F}_q$.
    \item If $C$ is a $\lambda$-constacyclic code over $\mathbb{F}_q$ and $r$ divides $(1+p^{e-k})$, then both $C^{\perp_k}$ and $\mathrm{hull}_k(C)$ are $\lambda$-constacyclic codes over $\mathbb{F}_q$.
\end{enumerate}
\end{lemma}

\begin{theorem}\label{Theorem21}\cite{LiuX}
    Let $C=\langle g(x)\rangle$ be a $\lambda$-constacyclic code of  dimension $l$ in $\mathbb{F}_q^n$. Let $h(x) = \sum_{i=0}^l h_ix^i$ be the parity-check polynomial of $C$. Then the $k$-Galois dual $C^{\perp_k}$ is generated by the polynomial
    \begin{equation}
        h^{\#}(x) = \sum_{i=0}^l h_0^{-p^{e-k}}h_i^{p^{e-k}}x^{l-i}.
    \end{equation}
Moreover, $\mathrm{hull}_k(C)$ is generated by $\mathrm{lcm}(g(x),h^{\#}(x))$ and $(\mathrm{hull}_k(C))^{\perp_k}$ is generated by $\gcd(g(x),h^{\#}(x))$.
\end{theorem}
From now on in this paper, for a given polynomial $f(x) = \sum_{i=0}^n f_ix^i \in \mathbb F_q[x]$, we will denote by $f^{\#}(x)$   the polynomial $\sum_{i=0}^n f_0^{-p^{e-k}}f_i^{p^{e-k}}x^{n-i}$.

\subsection{Constacyclic codes over $\mathcal A$}

A linear code $C$ of length $n$ over $\mathcal A$ is an $\mathcal A$-submodule of $\mathcal A^n$.
Since $\mathcal A = \sum_{S(\nu)\in \Hat{S}} \left(e_{S(\nu)}\mathbb{F}_q\right)$,  each element $a\in \mathcal A$ can be expressed as $$a = \sum_{S(\nu)\in \Hat{S}} a_{S(\nu)} e_{S(\nu)},$$ where $a_{S(\nu)} \in \mathbb{F}_q$ for all $S(\nu)\in \Hat{S}$. Also, for each $\rho = (\rho_0, \rho_1, \ldots, \rho_{n-1}) \in \mathcal A^n$, the projection of $\rho$ by $e_{S(\nu)}$ is given by $\rho_{S(\nu)} = (\rho_{0,S(\nu)}, \rho_{1,S(\nu)}, \ldots, \rho_{n-1,S(\nu)}) \in \mathbb{F}_q^n$, where $\rho_i = \sum_{S(\nu)\in \Hat{S}} \rho_{i,S(\nu)} e_{S(\nu)}$, $\rho_{i,S(\nu)} \in \mathbb{F}_q$ for $i = 0,1,\ldots,n-1$.
For  each   linear codes $C\subseteq \mathcal A^n$ and $S(\nu)\in\Hat{S}$, we define the projection code over $\mathbb F_q^n$ given by
\begin{equation}
    C_{S(\nu)} = \{ c_{S(\nu)} \in \mathbb{F}_q^n \mid c\in C \}.
\end{equation}
Thus, we have $C = \sum_{S(\nu)\in\Hat{S}} C_{S(\nu)} e_{S(\nu)}$. 

For any two elements $\rho = (\rho_0, \rho_1,\ldots, \rho_{n-1}), \rho^{\prime} = (\rho_0^{\prime}, \rho_1^{\prime},\ldots, \rho_{n-1}^{\prime}) \in\mathcal A^n$, we define the $k$-Galois inner product of $\rho$ and $\rho^{\prime}$ as
$
    \langle \rho, \rho^{\prime}\rangle_k = \sum_{i-0}^{n-1} \rho_i (\rho_i^{\prime})^{p^k}$.
and the $k$-Galois dual code $C^{\perp_k}$ of a linear code $C\subseteq \mathcal A^n$ as
$
    C^{\perp_k} = \{ c^{\prime}\in \mathcal A^n \mid \langle c^{\prime},c\rangle_k = 0 \text{ for all } c\in C\}$.  
Note that, since the basis for $\mathcal A$ is a set of primitive idempotents, it is easy to check that if $C = \sum_{S(\nu)\in\Hat{S}} C_{S(\nu)} e_{S(\nu)}$ then $$C^{\perp_k} = \sum_{S(\nu)\in\Hat{S}} C_{S(\nu)}^{\perp_k} e_{S(\nu)}.$$

For  $\lambda$  a unit in $\mathcal A$,    a linear code $C\subseteq \mathbb \mathcal A ^n$  is said to be a $\lambda$-constacyclic code if it can identify with an ideal in   $\frac{ \mathcal A [x]}{\langle x^n-\lambda\rangle}$. The following results in \cite{Yilmaz} on constacyclic codes over $\mathcal A$ will be required later in this work.

\begin{lemma}\label{Lemma3}
    An element $\lambda = \sum_{S(\nu)\in\Hat{S}} \lambda_{S(\nu)}e_{S(\nu)}\in \mathcal A$  is a unit in $\mathcal A$  if and only if for each $S(\nu)\in\Hat{S}$, $\lambda_{S(\nu)}$ is a unit in $\mathbb{F}_q$.
\end{lemma}

\begin{theorem}\label{Theorem22}
    Let $\lambda = \sum_{S(\nu)\in\Hat{S}} \lambda_{S(\nu)}e_{S(\nu)}$ be a unit in $\mathcal A$. Then $C$ is a $\lambda$-constacyclic code in $\mathcal A^n$  if and only if for each $S(\nu)\in\Hat{S}$, $C_{S(\nu)}$ is a $\lambda_{S(\nu)}$-constacyclic code of length $n$ over $\mathbb{F}_q$.
\end{theorem}

\begin{theorem}\label{Theorem23}
    Let $\lambda = \sum_{S(\nu)\in\Hat{S}} \lambda_{S(\nu)}e_{S(\nu)}$ be a unit in $\mathcal A$. Let $C$ be a $\lambda$-constacyclic code in $\mathcal A^n$  and $g(x)$ be the generator polynomial of $C$. Then $g(x)\mid x^n-\lambda$ over $\mathcal A$  and $g(x)$ is of the form $g(x) = \sum_{S(\nu)\in\Hat{S}} e_{S(\nu)}g_{S(\nu)}(x)$, where $g_{S(\nu)}(x)$ is the generator polynomial of $C_{S(\nu)}$ and $g_{S(\nu)}(x)\mid x^n-\lambda_{S(\nu)}$ over $\mathbb{F}_q$, for all $S(\nu)\in\Hat{S}$.
\end{theorem}

\subsection{Factorization of $x^n-\lambda$ in $\mathbb{F}_q[x]$}\label{Section2.3}

Let $n$ be a positive integer of the form $n=n^{\prime}p^{e^{\prime}}$ with $\gcd(n^{\prime},p) = 1$ and $e^{\prime}$ be a non-negative integer. Consider $\lambda \in \mathbb{F}_q^{\ast}$ and let $r=\mathrm{ord}(\lambda)$ be such that $r\mid (1+p^{e-k})$, where $k$ is a positive integer satisfying $0\leq k<e$. Let $l^{\prime}$ be the least multiple of $\mathrm{lcm}(e,e-k)$ such that $l=\frac{l^{\prime}}{e-k}$ is an even integer. Then, following \cite{Indibar} for $f(x)\in\mathbb{F}_q[x]$, we have $f^{l\#}(x) = u f(x)$, where $u$ is a unit in $\mathbb{F}_q$. Without loss of generality, we assume $a_1=1$ and let $a_1,a_2,\ldots,a_s$ be the complete list of distinct positive divisors of $l$. From \cite{Indibar}, we have the following factorization of $x^n-\lambda$ over $\mathbb{F}_q$.
\begin{align}\label{Equation1}
    x^n-\lambda =&~ \prod_{t=1}^s \prod_{j\in D_t} \prod_{i=1}^{\beta_{t}(j)} \Big\{f_{ij}(x)f_{ij}^{\#}(x)\cdots f_{ij}^{(a_t-1)\#}(x)\Big\}^{p^{e^{\prime}}}, \hbox{ where } \beta_t(j) = \frac{\phi(j)}{a_t\phi(r)\mathrm{ord}_j(q)},
\end{align}
 and $D_t = A\cap B_{a_t}$ for $t=1,2,\ldots,s$.
For $A$ and $B_i$ defined as follows
\begin{align*}
    A = \Big\{j\in\mathbb{N} \mid \gcd(j,q) = 1, j \text{ divides } n^{\prime}r \text{ and } \gcd\big(\frac{n^{\prime}r}{j},r\big) = 1\Big\},
\end{align*}
\begin{align*}
    B_i =&~ \Big\{j\in\mathbb{N} \mid i \text{ is the least positive integer with } j \mid (p^{el_1}+(-1)^{i-1}p^{i(e-k)}) \text{ for at least one}\\
    &~~\text{ non-negative integer } l_1 \Big\}.
\end{align*}

\section{Galois hulls of constacyclic codes over $\mathcal A$}\label{sec.3}

In this section, we find the form of the generator of the $k$-Galois dual as well as the $k$-Galois hull of a $\lambda$-constacyclic code of length $n$ over $\mathcal A$. We also derive a formula for the $q$-dimension of the $k$-Galois hull of a $\lambda$-constacyclic code. Further, we present a condition for a $\lambda$-constacyclic code over $\mathcal A$ to be a $k$-Galois LCD code. 

\begin{theorem}\label{Theorem31}
    Let $\lambda$ be a unit in $\mathcal A$. If $C = \sum_{S(\nu)\in\Hat{S}} C_{S(\nu)}e_{S(\nu)}$ is a $\lambda$-constacyclic code of length $n$ over $\mathcal A$ then $C^{\perp_k}$ is a $(\lambda^{-p^{e-k}})$-constacyclic code of length $n$ over $\mathcal A$.
\end{theorem}

\begin{proof}
    The proof follows from Proposition $3$ of \cite{Yilmaz} and Lemma \ref{Lemma1}.\qed
\end{proof}

\begin{theorem}\label{Theorem32}
    Let $\lambda = \sum_{S(\nu)\in\Hat{S}} \lambda_{S(\nu)}e_{S(\nu)}$ be a unit in $\mathcal A$ and let $C = \bigoplus_{S(\nu)\in\Hat{S}} C_{S(\nu)}e_{S(\nu)}$ be a $\lambda$-constacyclic code of length $n$ over $\mathcal A$ with  $g(x) = \sum_{S(\nu)\in\Hat{S}} e_{S(\nu)}g_{S(\nu)}(x)$ being its generator polynomial, where $g_{S(\nu)}(x)$ is the generator polynomial of $C_{S(\nu)}$ for $S(\nu)\in\Hat{S}$. Then $C^{\perp_k}$ is generated by the polynomial $\Tilde{h}(x) = \sum_{S(\nu)\in\Hat{S}} e_{S(\nu)}h_{S(\nu)}^{\#}(x)$, where $h_{S(\nu)}(x) = \frac{x^n-\lambda_{S(\nu)}}{g_{S(\nu)}(x)}$ for $S(\nu)\in\Hat{S}$.
\end{theorem}

\begin{proof}
    We know that $C^{\perp_k} = \bigoplus_{S(\nu)\in\Hat{S}} C_{S(\nu)}^{\perp_k}e_{S(\nu)}$. Now, the result follows from Theorem \ref{Theorem21} and Theorem \ref{Theorem23}.\qed
\end{proof}

\begin{lemma}\label{Lemma3.1}
    Let $\lambda \in \mathbb{F}_q$ be a unit and $C = \langle g(x)\rangle$ be a $\lambda$-constacyclic code of length $n$ over $\mathbb{F}_q$. Then $C$ contains its $k$-Galois dual if and only if $$x^n-\lambda \equiv 0\pmod{g(x)g^{(l-1)\#}(x)},$$ where $l$ is the same integer mentioned in Section \ref{Section2.3}.
\end{lemma}

\begin{proof}
    First, let $C$ be a $\lambda$-constacyclic code containing its $k$-Galois dual, i.e., $C^{\perp_k}\subseteq C$. From Theorem \ref{Theorem21}, we know that $C^{\perp_k}$ is generated by $h^{\#}(x)$, where $h(x)$ is the parity-check polynomial of $C$. Thus, $g(x)\mid h^{\#}(x)$, and hence $h^{\#}(x) = g(x)g_1(x)$ for some $g_1(x)\in \mathbb{F}_q[x]$. Now,
    \begin{align*}
        & h^{\#}(x) = g(x)g_1(x)\\
        \Rightarrow &~ h^{l\#}(x) = g^{(l-1)\#}(x)g_1^{(l-1)\#}(x)\\
        \Rightarrow &~ u h(x) = g^{(l-1)\#}(x)g_1^{(l-1)\#}(x)\quad\quad\quad (\mathrm{for~ some~ } u\in\mathbb{F}_q^*)\\
        \Rightarrow &~ x^n-1 = g(x) h(x) = u^{-1} g(x)g^{(l-1)\#}(x)g_1^{(l-1)\#}(x)\\
        \Rightarrow &~ x^n-\lambda \equiv 0\pmod{g(x)g^{(l-1)\#}(x)}.
    \end{align*}
    Conversely, let $x^n-\lambda \equiv 0\pmod{g(x)g^{(l-1)\#}(x)}$. Then $g(x)g^{(l-1)\#}(x)$ divides $x^n-\lambda$, i.e., $g(x)g^{(l-1)\#}(x)$ divides $g(x)h(x)$. This implies $g^{(l-1)\#}(x)$ divides $h(x)$, and hence $h(x) = g^{(l-1)\#}(x) g_2(x)$ for some $g_2(x)\in\mathbb{F}_q[x]$. Now,
    \begin{align*}
        & h(x) = g^{(l-1)\#}(x) g_2(x)\\
        \Rightarrow &~ h^{\#}(x) = g^{l\#}(x) g_2^{\#}(x)\\
        \Rightarrow &~ h^{\#}(x) = u g(x)g_2^{\#}(x),
    \end{align*}
    where $u$ is a unit in $\mathbb{F}_q$. Thus, $g(x)\mid h^{\#}(x)$, and hence $\langle h^{\#}(x)\rangle \subseteq \langle g(x)\rangle$. Therefore, $C$ contains its $k$-Galois dual.\qed
\end{proof}

\begin{theorem}\label{Theorem321}
    Let $\lambda = \sum_{S(\nu)\in\Hat{S}} \lambda_{S(\nu)}e_{S(\nu)}$ be a unit in $\mathcal A$ and $C = \bigoplus_{S(\nu)\in\Hat{S}} C_{S(\nu)} e_{S(\nu)}$ be a $\lambda$-constacyclic code of length $n$ over $\mathcal A$ with generator polynomial $g(x) = \sum_{S(\nu)\in\Hat{S}} e_{S(\nu)} g_{S(\nu)}(x)$. Then $C$ contains its $k$-Galois dual if and only if $$x^n-\lambda_{S(\nu)} \equiv 0\pmod{g_{S(\nu)}(x)g_{S(\nu)}^{(l-1)\#}(x)}$$ for all $S(\nu)\in\Hat{S}$.
\end{theorem}

\begin{proof}
   Note  that $C^{\perp_k} = \bigoplus_{S(\nu)\in\Hat{S}} C_{S(\nu)}^{\perp_k} e_{S(\nu)}$, thus $C^{\perp_k}\subseteq C$ if and only if $C_{S(\nu)}^{\perp_k}\subseteq C_{S(\nu)}$ for all $S(\nu)\in\Hat{S}$. Now, the proof follows from Lemma \ref{Lemma3.1}.\qed
\end{proof}

\begin{corollary}
     Let $\lambda = \sum_{S(\nu)\in\Hat{S}} \lambda_{S(\nu)}e_{S(\nu)}$ be a unit in $\mathcal A$ and $C = \bigoplus_{S(\nu)\in\Hat{S}} C_{S(\nu)} e_{S(\nu)}$ be a $\lambda$-constacyclic code of length $n$ over $\mathcal A$. Then the following statements are true.
     \begin{enumerate}
         \item $C$ is a $k$-Galois self-dual code over $\mathcal A$ if and only if for each $S(\nu)\in\Hat{S}$, $C_{S(\nu)}$ is a $k$-Galois self-dual code over $\mathbb{F}_q$.
         \item $C$ is a $k$-Galois LCD code over $\mathcal A$ if and only if for each $S(\nu)\in\Hat{S}$, $C_{S(\nu)}$ is a $k$-Galois LCD code over $\mathbb{F}_q$.
     \end{enumerate}
\end{corollary}

\begin{lemma}\label{Theorem322}
    Let $C_1$ and $C_2$ be two $k$-Galois self-dual codes of lengths $n_1$ and $n_2$ over $\mathbb{F}_q$. Then $C_1\times C_2$, the direct product of $C_1$ and $C_2$ is a $k$-Galois self-dual code of length $n_1+n_2$ over $\mathbb{F}_q$.
    
\end{lemma}
\begin{proof}
    Take $(a_1,a_2), (b_1,b_2) \in C_1\times C_2$. Then $a_i, b_i\in C_i$ for $i=1,2$. Since $C_1$ and $C_2$ are $k$-Galois self-dual codes, we have $\langle a_1, b_1 \rangle_k = 0 = \langle a_2, b_2 \rangle_k$. Hence,
    $
        \langle (a_1,a_2), (b_1,b_2) \rangle_k = \langle a_1, b_1 \rangle_k + \langle a_2, b_2 \rangle_k = 0.
$    This implies $C_1\times C_2 \subseteq (C_1\times C_2)^{\perp_k}$. 

On the other side, since $C_i$ is a $k$-Galois self-dual code, we have $\mid C_i\mid~ = q^{\frac{n_i}{2}} = ~\mid C_i^{\perp_k}\mid$ for $i=1,2$. Thus, $\mid C_1\times C_2 \mid~ = ~\mid C_1 \mid \times \mid C_2 \mid~ = q^{\frac{n_1}{2}}\times q^{\frac{n_2}{2}} = q^{\frac{n_1+n_2}{2}}$, and hence $\mid (C_1\times C_2)^{\perp_k} \mid~ = \frac{q^{n_1+n_2}}{\mid C_1\times C_2 \mid} = q^{\frac{n_1+n_2}{2}} = ~\mid C_1\times C_2 \mid$. Therefore, $C_1\times C_2 = (C_1\times C_2)^{\perp_k}$, i.e., $C_1\times C_2$ is a $k$-Galois self-dual code.\qed
\end{proof}

\begin{corollary}
    If there is a $k$-Galois self-dual code of length $2$ over $\mathbb{F}_q$, then there exists a $k$-Galois self-dual code of all even lengths over $\mathbb{F}_q$.
\end{corollary}

\begin{proof}
    Take an even integer $n=2m$ for some positive integer $m$. Let $C$ be a $k$-Galois self-dual code of length $2$ over $\mathbb{F}_q$. Let us denote $D = \overbrace{C\times \cdots \times C}^{m \mathrm{ times}}$. Then from Theorem \ref{Theorem322}, $D$ is a $k$-Galois self-dual code of length $2m$.\qed
\end{proof}
Thus, we get the following corollary directly from the above results.
\begin{corollary}
    If a $k$-Galois self-dual code of length $2$ over $\mathbb{F}_q$ exists, then there exist $k$-Galois self-dual codes of all even lengths over $\mathcal A$.
\end{corollary}
 
\begin{theorem}\label{Theorem33}
        Let $\lambda = \sum_{S(\nu)\in\Hat{S}} \lambda_{S(\nu)}e_{S(\nu)}$ be a unit in $\mathcal A$ and let $C = \bigoplus_{S(\nu)\in\Hat{S}} C_{S(\nu)}e_{S(\nu)}$ be a $\lambda$-constacyclic code of length $n$ over $\mathcal A$ with  $g(x) = \sum_{S(\nu)\in\Hat{S}} e_{S(\nu)}g_{S(\nu)}(x)$ being its generator polynomial, where $g_{S(\nu)}(x)$ is the generator polynomial of $C_{S(\nu)}$ for $S(\nu)\in\Hat{S}$. Then $\mathrm{hull}_k(C)$  is given by $$\mathrm{hull}_k(C) = \bigoplus_{S(\nu)\in\Hat{S}} \mathrm{hull}_k(C_{S(\nu)})$$ and it has as  generator polynomial $$G(x) = \sum_{S(\nu)\in\Hat{S}} e_{S(\nu)}\mathrm{lcm}(g_{S(\nu)}(x),h_{S(\nu)}^{\#}(x)),$$ where $h_{S(\nu)}(x) = \frac{x^n-\lambda_{S(\nu)}}{g_{S(\nu)}(x)}$ for $S(\nu)\in\Hat{S}$.
\end{theorem}

\begin{proof}
    The proof follows from Theorem \ref{Theorem23} and Theorem \ref{Theorem32}.\qed
\end{proof}

Let $\lambda = \sum_{S(\nu)\in\Hat{S}} \lambda_{S(\nu)}e_{S(\nu)}$ be a unit in $\mathcal A$. Then from Lemma \ref{Lemma3}, we know that $\lambda_{S(\nu)}$ is a unit in $\mathbb{F}_q$ for each $S(\nu)\in\Hat{S}$. Let $\mathrm{ord}(\lambda_{S(\nu)}) = r_{S(\nu)}$ and $\mathcal{A}_{S(\nu)}$ be the set
\begin{align*}
    \mathcal{A}_{S(\nu)} = \Big\{j\in\mathbb{N} : \gcd(j,q) = 1,~ j \text{ divides } n^{\prime}r_{S(\nu)} \text{ and } \gcd\big(\frac{n^{\prime}r_{S(\nu)}}{j}, r_{S(\nu)}\big) = 1\Big\}, \quad \mathrm{for } S(\nu)\in\Hat{S}.
\end{align*}
Then similar to Equation~(\ref{Equation1}), $x^n-\lambda_{S(\nu)}$ can be factorized over $\mathbb{F}_q$ as
\begin{align}\label{Equation13}
    x^n-\lambda_{S(\nu)} =&~ \prod_{t=1}^s \prod_{j\in D_{S(\nu) t}} \prod_{i=1}^{\beta_{S(\nu) t}(j)} \Big\{f_{ij}(x)f_{ij}^{\#}(x)\cdots f_{ij}^{(a_t-1)\#}(x)\Big\}^{p^{e^{\prime}}},
\end{align}
where $$\beta_{S(\nu) t}(j) = \frac{\phi(j)}{a_t\phi(r_{S(\nu)})\mathrm{ord}_j(q)} \hbox{ and } D_{S(\nu) t} = \mathcal{A}_{S(\nu)}\cap B_{a_t}     \hbox{ for }1\leq t\leq s,\, (\nu)\in\Hat{S}.$$
From the factorization in Equation~(\ref{Equation13}), we have
\begin{align*}
\sum_{t=1}^s \sum_{j\in D_{S(\nu) t}} \sum_{i=1}^{\beta_{S(\nu) t}(j)} \Big\{\deg\big(f_{ij}(x)f_{ij}^{\#}(x)\cdots f_{ij}^{(a_t-1)\#}(x)\big)\Big\} = n^{\prime}.
\end{align*}
If
$\Delta_{S(\nu) t} = \sum_{j\in D_{S(\nu) t}} \sum_{i=1}^{\beta_{S(\nu) t}(j)} \deg(f_{ij}(x))$,
then
\begin{equation}\label{Equation32}
    \sum_{t=1}^s a_t\Delta_{S(\nu) t} = n^{\prime}.
\end{equation}

Now, we derive an expression for $\Delta_{S(\nu) t}$, which will be used later.

\begin{lemma}\label{Lemma5} With the notation above, 
    $\Delta_{S(\nu) t} = \displaystyle\sum_{j\in D_{S(\nu) t}} \frac{\phi(j)}{a_t\phi(r_{S(\nu)})}$.
\end{lemma}

\begin{proof}
    We know, $\displaystyle\sum_{t=1}^s a_t\Delta_{S(\nu) t} = n^{\prime}$. Also, combining Lemma $3$ and Lemma $4$ in \cite{Indibar}, we have $$\displaystyle\sum_{t=1}^s\sum_{j\in D_{S(\nu) t}} \frac{\phi(j)}{\phi(r_{S(\nu)})} = n^{\prime}.$$ Thus,
    $$\sum_{t=1}^s a_t\Delta_{S(\nu) t} = \sum_{t=1}^s\sum_{j\in D_{S(\nu) t}} \frac{\phi(j)}{\phi(r_{S(\nu)})}
        \Rightarrow a_t\Delta_{S(\nu) t} = \sum_{j\in D_{S(\nu) t}} \frac{\phi(j)}{\phi(r_{S(\nu)})}
        \Rightarrow  \Delta_{S(\nu) t} = \displaystyle\sum_{j\in D_{S(\nu) t}} \frac{\phi(j)}{a_t\phi(r_{S(\nu)})}.
    $$
    This completes the proof.\qed
\end{proof}

\begin{lemma}\label{Lemma9}
    Let $\lambda = \sum_{S(\nu)\in\Hat{S}} \lambda_{S(\nu)}e_{S(\nu)}$ be a unit in $\mathcal A$. Then $\mathrm{ord}(\lambda_{S(\nu)})$ divides $\mathrm{ord}(\lambda)$ for each $S(\nu)\in\Hat{S}$.
\end{lemma}

\begin{proof}
    Let $\mathrm{ord}(\lambda) = r$ and $\mathrm{ord}(\lambda_{S(\nu)}) = r_{S(\nu)}$. Hence, since there is a decomposition in primitive idempotents of $\mathcal A$ one has
    \begin{align*}
        &~\lambda^r = 1\\
        \Rightarrow &~ \sum_{S(\nu)\in\Hat(S)}\lambda_{S(\nu)}^r e_{S(\nu)} = 1\\
        \Rightarrow &~ \sum_{S(\nu)\in\Hat(S)}\lambda_{S(\nu)}^r e_{S(\nu)} = \sum_{S(\nu)\in\Hat(S)} e_{S(\nu)}\\
        \Rightarrow &~ \lambda_{S(\nu)}^r = 1\quad \forall~ S(\nu)\in\Hat{S},
    \end{align*}
    and so we have $r_{S(\nu)}\mid r$ for each $S(\nu)\in\Hat{S}$.\qed
\end{proof}

\begin{theorem}\label{Theorem13}
	Let $n$ be a positive integer of the form $n^{\prime}p^{e^{\prime}}$, where $\gcd(n^{\prime},p) = 1$, $e^{\prime}$ being a non-negative integer and $\lambda = \sum_{S(\nu)\in\Hat{S}} \lambda_{S(\nu)}e_{S(\nu)}$ be a unit in $\mathcal A$ such that $\gcd(\mathrm{ord}(\lambda),n^{\prime}) = 1$ and $\mathrm{ord}(\lambda)$ divides $1+p^{e-k}$, where $0\leq k<e$. Then the $q$-dimension of the $k$-Galois hull of a $\lambda$-constacyclic code $C$ of length $n$ over $\mathcal A$ is of the form
 \begin{align*}
     \sum_{S(\nu)\in\Hat{S}} \sum_{t = 1} ^s \sum_{j\in D_{S(\nu) t}} \mathrm{ord}_j(q)\cdot b_{S(\nu) tj},
 \end{align*}
 where $b_{S(\nu) tj} = \displaystyle\sum_{i = 1} ^{\beta_{S(\nu) t}(j)} b_{S(\nu) tij},~ b_{S(\nu) tij} = a_t p^{e^{\prime}} - \big[\max\{u_{S(\nu) tij,0},{p^{e^{\prime}}-u_{S(\nu) tij,a_t-1}}\} + \max\{u_{S(\nu) tij,1},\\
 {p^{e^{\prime}}-u_{S(\nu) tij,0}}\}+\cdots+\max\{u_{S(\nu) tij,a_t-1},{p^{e^{\prime}}-u_{S(\nu) tij,a_t-2}}\}\big]$ and $0\leq b_{S(\nu) tj}\leq \beta_{S(\nu) t}(j)\cdot \lfloor \frac{a_t p^{e^{\prime}}}{2}\rfloor$ for $1\leq t\leq s$, $S(\nu)\in\Hat{S}$.
\end{theorem}

\begin{proof}
    Let $\mathrm{ord}(\lambda) = r$ and $\mathrm{ord}(\lambda_{S(\nu)}) = r_{S(\nu)}$. It is given that $r\mid (1+p^{e-k})$ and hence from Lemma \ref{Lemma1}, we get that $\mathrm{hull}_k(C)$ is a $\lambda$-constacyclic code over $\mathcal A$. Also, we know that $C = \bigoplus_{S(\nu)\in\Hat{S}} C_{S(\nu)} e_{S(\nu)}$, where $C_{S(\nu)}$ are $\lambda_{S(\nu)}$-constacyclic codes over $\mathbb{F}_q$. Then from Lemma \ref{Lemma9} and Lemma \ref{Lemma1}, we get that for each $S(\nu)\in\Hat{S}$, $\mathrm{hull}_k(C_{S(\nu)})$ is a $\lambda_{S(\nu)}$-constacyclic code over $\mathbb{F}_q$. Then from Theorem \ref{Theorem33}, the $q$-dimension of the $k$-Galois hull of the $\lambda$-constacyclic code $C$ is given by
    \begin{align}\label{Equation14}
        \mathrm{dim}(\mathrm{hull}_k(C))
        =&~ \sum_{S(\nu)\in\Hat{S}}n - \sum_{S(\nu)\in\Hat{S}} \deg(\mathrm{lcm}(g_{S(\nu)}(x),h_{S(\nu)}^{\#}(x)))\nonumber\\
         =&~ \sum_{S(\nu)\in\Hat{S}} \big\{n-\deg(\mathrm{lcm}(g_{S(\nu)}(x),h_{S(\nu)}^{\#}(x)))\big\},
    \end{align}
    where $g_{S(\nu)}(x)$ and $h_{S(\nu)}(x)$ are respectively the generator polynomial and the parity-check polynomial of $C_{S(\nu)}$. Now, from Equation~(\ref{Equation13}), for each $S(\nu)\in\Hat{S}$, the generator polynomial of $C_{S(\nu)}$ is given by
    \begin{align*}
        g_{S(\nu)}(x) = \prod_{t=1}^s \prod_{j\in D_{S(\nu) t}} \prod_{i=1}^{\beta_{S(\nu) t}(j)} (f_{ij}(x))^{u_{S(\nu) tij,0}}(f_{ij}^{\#}(x))^{u_{S(\nu) tij,1}}\cdots (f_{ij}^{(a_t-1)\#}(x))^{u_{S(\nu) tij,a_t-1}},
    \end{align*}
    where $0\leq u_{S(\nu) tij,\delta}\leq p^{e^{\prime}}$ for $0\leq \delta\leq a_t-1$.
    The parity-check polynomial of $C_{S(\nu)}$ is given by
    \begin{align*}
        h_{S(\nu)}(x) =&~ \prod_{t=1}^s \prod_{j\in D_{S(\nu) t}} \prod_{i=1}^{\beta_{S(\nu) t}(j)}\Big[ (f_{ij}(x))^{p^{e^{\prime}} - u_{S(\nu) tij,0}}(f_{ij}^{\#}(x))^{p^{e^{\prime}} - u_{S(\nu) tij,1}}\times\cdots\\
        &~~~~~~~~~~~~~~~~~~~~~~~~~~~~~\cdots\times (f_{ij}^{(a_t-1)\#}(x))^{p^{e^{\prime}} - u_{S(\nu) tij,a_t-1}}\Big].
    \end{align*}
    Then
    \begin{align*}
        h_{S(\nu)}^{\#}(x) =&~ \prod_{t=1}^s \prod_{j\in D_{S(\nu) t}} \prod_{i=1}^{\beta_{S(\nu) t}(j)} \Big[(f_{ij}(x))^{p^{e^{\prime}} - u_{S(\nu) tij,a_t-1}}(f_{ij}^{\#}(x))^{p^{e^{\prime}} - u_{S(\nu) tij,0}}\times\cdots\\
        &~~~~~~~~~~~~~~~~~~~~~~~~~~~~~\cdots\times(f_{ij}^{(a_t-1)\#}(x))^{p^{e^{\prime}} - u_{S(\nu) tij,a_t-2}}\Big]
    \end{align*}
    and
    \begin{align*}
        &~ \mathrm{lcm}(g_{S(\nu)}(x),h_{S(\nu)}^{\#}(x)) \\
        =&~ \prod_{t=1}^s \prod_{j\in D_{S(\nu) t}} \prod_{i=1}^{\beta_{S(\nu) t}(j)} \Big[(f_{ij}(x))^{\max\{u_{S(\nu) tij,0},p^{e^{\prime}} - u_{S(\nu) tij,a_t-1}\}} (f_{ij}(x))^{\max\{u_{S(\nu) tij,1},p^{e^{\prime}} - u_{S(\nu) tij,0}\}}\times\cdots\\
        &~~~~~~~~~~~~~~~~~~~~~~~~~~~~~\cdots\times (f_{ij}^{(a_t-1)\#}(x))^{\max\{u_{S(\nu) tij,a_t-1},p^{e^{\prime}} - u_{S(\nu) tij,a_t-2}\}}\Big].
    \end{align*}    
    Again, from the proof of Theorem $3$ in \cite{Indibar}, for each $S(\nu)\in\Hat{S}$, we have 
    \begin{align*}
        &~ n-\deg(\mathrm{lcm}(g_{S(\nu)}(x),h_{S(\nu)}^{\#}(x)))\nonumber\\
        =&~ \sum_{t = 1} ^s \sum_{j\in D_{S(\nu) t}} a_t \cdot \mathrm{ord}_j(q) \cdot \sum_{i = 1} ^{\beta_{S(\nu) t}(j)} p^{e^{\prime}}\nonumber\\
    &~ - \sum_{t = 1} ^s \sum_{j\in D_{S(\nu) t}} \sum_{i = 1} ^{\beta_{S(\nu) t}(j)}\mathrm{ord}_j(q)\Big[ \max\{u_{S(\nu) tij,0},{p^{e^{\prime}}-u_{S(\nu) tij,a_t-1}}\} +\max\{u_{S(\nu) tij,1},{p^{e^{\prime}}-u_{S(\nu) tij,0}}\} +\nonumber\\
    & ~~~~~~~~~~~~~~~~~~~~~~~~~~~~~~~~~~~~~~~~~~~\cdots + \max\{u_{S(\nu) tij,a_t-1},{p^{e^{\prime}}-u_{S(\nu) tij,a_t-2}}\}\Big]\nonumber\\
     \end{align*}
        \begin{align}\label{Equation15}   
    =&~ \sum_{t = 1} ^s \sum_{j\in D_{S(\nu) t}} \mathrm{ord}_j(q) \cdot \sum_{i = 1} ^{\beta_{S(\nu) t}(j)} b_{S(\nu) tij}\nonumber\\
    =&~ \sum_{t = 1} ^s \sum_{j\in D_{S(\nu) t}} \mathrm{ord}_j(q)\cdot b_{S(\nu) tj},
    \end{align}
    where $b_{S(\nu) tj} = \displaystyle\sum_{i = 1} ^{\beta_{\alpha t}(j)} b_{S(\nu) tij}$, $b_{S(\nu) tij} = a_t p^{e^{\prime}} - \big[\max\{u_{S(\nu) tij,0},{p^{e^{\prime}}-u_{S(\nu) tij,a_t-1}}\} + \max\{u_{S(\nu) tij,1},\\
    {p^{e^{\prime}}-u_{S(\nu) tij,0}}\}+\cdots+\max\{u_{S(\nu) tij,a_t-1},{p^{e^{\prime}}-u_{S(\nu) tij,a_t-2}}\}\big]$ and $0\leq b_{S(\nu) tj}\leq \beta_{S(\nu) t}(j)\cdot \lfloor \frac{a_t p^{e^{\prime}}}{2}\rfloor$ for $1\leq t\leq s$.
    
    Now, combining Equations (\ref{Equation14}) and (\ref{Equation15}), we get the required form of the $q$-dimension of the $k$-Galois hull of a $\lambda$-constacyclic code $C$ over $\mathcal A$.\qed
\end{proof}

The next result can be obtained directly from the above theorem when $q$ and $n$ are relative prime.

\begin{corollary}\label{Corollary4}
Let $n$ be a positive integer relatively prime to $q$ and $\lambda = \sum_{S(\nu)\in\Hat{S}} \lambda_{S(\nu)}e_{S(\nu)}$ be a unit in $\mathcal A$ such that $\gcd(\mathrm{ord}(\lambda),n) = 1$ and $\mathrm{ord}(\lambda)$ divides $1+p^{e-k}$, where $0\leq k<e$. Then the $q$-dimension of the $k$-Galois hull of a $\lambda$-constacyclic code $C$ of length $n$ over $\mathcal A$ is of the form
 \begin{align*}
     \sum_{S(\nu)\in\Hat{S}} \sum_{t = 2} ^s \sum_{j\in D_{S(\nu) t}} \mathrm{ord}_j(q)\cdot b_{S(\nu) tj},
 \end{align*}
 where $0\leq b_{S(\nu) tj}\leq \beta_{S(\nu) t}(j)\cdot \lfloor \frac{a_t}{2}\rfloor$ for $2\leq t\leq s$.
\end{corollary}

\begin{proof}
    We know that $a_1 = 1$ and thus for each $S(\nu)\in\Hat{S}$, we have
    $ 0\leq b_{S(\nu) 1j}\leq \beta_{S(\nu) 1}(j)\cdot \Big\lfloor \frac{a_1}{2}\Big\rfloor $, and hence $  b_{S(\nu) 1j} = 0$.
    Now, the proof follows directly from Theorem \ref{Theorem13}.\qed
\end{proof}

We define $\mathcal{D}_k(n,q,\lambda)$  as the set of all $q$-dimensions of $k$-Galois hulls of $\lambda$-constacyclic codes of length $n$ over $\mathcal A$. From Section \ref{Section2.1}, we recall that the number of equivalence classes $S(\nu)$ in $\Hat{S}$ is denoted by the notation $\mathcal{N}$. Now, we have the following corollary on the cardinality of $\mathcal{D}_k(n,q,\lambda)$.

\begin{corollary}
    Let $n$ be a positive integer of the form $n^{\prime}p^{e^{\prime}}$, where $\gcd(n^{\prime},p) = 1$, $e^{\prime}$ being a non-negative integer and $\lambda = \sum_{S(\nu)\in\Hat{S}} \lambda_{S(\nu)}e_{S(\nu)}$ be a unit in $\mathcal A$ such that $\gcd(\mathrm{ord}(\lambda),n^{\prime}) = 1$ and $\mathrm{ord}(\lambda)$ divides $1+p^{e-k}$, where $0\leq k<e$.
    \begin{enumerate}
    \item  If $p$ is odd and $n^{\prime} = 1$, then $\mid \mathcal{D}_k(n,q,\lambda)\mid~ = \frac{\mathcal{N}(n-1)}{2}+1$.
    \item  If $p=2$, $e$ is even, and $\mathrm{ord}(\lambda)n^{\prime}\mid (2^e-1)$, then $\mid \mathcal{D}_{\frac{e}{2}}(n,2^e,\lambda)\mid~ = \frac{\mathcal{N} n}{2}+1$.
    \end{enumerate}
\end{corollary}

\begin{proof}
    We know that $\mathcal{D}_k(n,q,\lambda)\subseteq\{0,1,2,\ldots,\lfloor\frac{n}{2}\rfloor\}$. Let $\mathrm{ord}(\lambda) = r$ and $\mathrm{ord}(\lambda_{S(\nu)}) = r_{S(\nu)}$ for each $S(\nu)\in\Hat{S}$.
    \begin{enumerate}
    \item  Here, $n$ is odd and so $\mathcal{D}_k(n,q,\lambda)\subseteq\{0,1,2,\ldots,\frac{n-1}{2}\}$. Also, we have
    \begin{align*}
        \mathcal{A}_{S(\nu)} = \Big\{j\in\mathbb{N} : \gcd(j,q) = 1,~ j \text{ divides } r_{S(\nu)} \text{ and } \gcd\big(\frac{r_{S(\nu)}}{j}, r_{S(\nu)}\big) = 1\Big\} = \Big\{r_{S(\nu)}\Big\},
    \end{align*}
    and recall the set
    \begin{align*}
    B_i =&~ \Big\{j\in\mathbb{N} : i \text{ is the least positive integer with } j \mid (p^{el_1}+(-1)^{i-1}p^{i(e-k)}) \text{ for at least one}\\
    &~~\text{ non-negative integer } l_1 \Big\}.
\end{align*}
It is given that $r$ divides $1+p^{e-k}$, and combining with Lemma \ref{Lemma9}, we thus have $r_{S(\nu)}$ divides $1+p^{e-k}$ for each $S(\nu)\in\Hat{S}$. This implies that $r_{S(\nu)} \in B_1$ for each $S(\nu)\in\Hat{S}$ and hence $D_{S(\nu)1} = \{r_{S(\nu)}\}$, and $D_{S(\nu)t} = \emptyset$ for all $S(\nu)\in\Hat{S}$ and $t = 2,3,4,\ldots,s$. Therefore, from Theorem \ref{Theorem13}, the $q$-dimension of the $k$-Galois hull of a $\lambda$-constacyclic code of length $n$ over $\mathcal A$ is given by
\begin{align*}
     \sum_{S(\nu)\in\Hat{S}}  \mathrm{ord}_{r_{S(\nu)}}(q)\cdot b_{S(\nu) 1 r_{S(\nu)}},
 \end{align*}
 where $b_{S(\nu) 1 r_{S(\nu)}} = \displaystyle\sum_{i = 1} ^{\beta_{S(\nu) 1}(r_{S(\nu)})} b_{S(\nu) 1i r_{S(\nu)}}$, and $0\leq b_{S(\nu) 1 r_{S(\nu)}}\leq \beta_{S(\nu) 1}(r_{S(\nu)})\cdot \lfloor \frac{p^{e^{\prime}}}{2}\rfloor$ for all $S(\nu)\in\Hat{S}$. We know that $\lambda_{S(\nu)}\in \mathbb{F}_q^{\ast}$, which implies that $r_{S(\nu)}$ divides $q-1$. Hence, $\mathrm{ord}_{r_{S(\nu)}}(q) = 1$. Also, $$\beta_{S(\nu) 1}(r_{S(\nu)}) = \frac{\phi(r_{S(\nu)})}{a_1 \cdot \phi(r_{S(\nu)}) \cdot \mathrm{ord}_{r_{S(\nu)}}(q)} = 1.$$ Thus, $$0\leq b_{S(\nu) 1 r_{S(\nu)}}\leq \frac{p^{e^{\prime}}-1}{2},$$ i.e., $0\leq b_{S(\nu) 1 r_{S(\nu)}}\leq \frac{n-1}{2}$. So, the set $\{0,1,2,\ldots,\mathcal{N}(\frac{n-1}{2})\}$ represents all the possible elements of the set $\mathcal{D}_k(n,q,\lambda)$ and therefore, $$\mid \mathcal{D}_k(n,q,\lambda)\mid~ = \frac{\mathcal{N}(n-1)}{2}+1.$$
 \item  It is given that $rn^{\prime}\mid (2^e-1)$ and since $r_{S(\nu)}\mid r$, we have $r_{S(\nu)}n^{\prime}\mid (2^e-1)$. Also, $j\in D_{S(\nu) t}$ implies that $j\in \mathcal{A}_{S(\nu)}$. Thus, for all $j\in D_{S(\nu) t}$, we have $j\mid (2^e-1)$. Hence, $\mathrm{ord}_j(q) = 1$ for all $j\in D_{S(\nu) t}$. Further, we have $e$ is an even integer and $k = \frac{e}{2}$. This implies $l=2$ and hence $s=2$ with $a_1 = 1$ and $a_2 = 2$. Now, from Theorem \ref{Theorem13}, we have the $q$-dimension of the $k$-Galois hull of a $\lambda$-constacyclic code of length $n$ over $\mathcal A$ as
\begin{align*}
     \sum_{S(\nu)\in\Hat{S}} \sum_{t = 1} ^2 \sum_{j\in D_{S(\nu) t}} b_{S(\nu) tj},
 \end{align*}
 where $0\leq b_{S(\nu) tj }\leq \beta_{S(\nu) t}(j)\cdot \lfloor \frac{a_t p^{e^{\prime}}}{2}\rfloor$ for all $j\in D_{S(\nu) t}$,  $t=1,2$, and $S(\nu)\in\Hat{S}$. Now,
 \begin{align*}
     \beta_{S(\nu) t}(j) \lfloor \frac{a_t p^{e^{\prime}}}{2}\rfloor =&~ \sum_{S(\nu)\in\Hat{S}} \sum_{t = 1} ^2 \sum_{j\in D_{S(\nu) t}} \frac{\phi(j)}{a_t \phi(r_{S(\nu)})} \Big\lfloor \frac{a_t p^{e^{\prime}}}{2}\Big\rfloor\\
     =&~ \sum_{S(\nu)\in\Hat{S}} \sum_{t = 1} ^2 \Delta_{S(\nu) t} \Big\lfloor \frac{a_t p^{e^{\prime}}}{2}\Big\rfloor \quad\quad\quad\quad[\text{From Lemma } \ref{Lemma5}]\\
     =&~ \sum_{S(\nu)\in\Hat{S}} \{ \Delta_{S(\nu) 1}\cdot\frac{ p^{e^{\prime}}}{2} + \Delta_{S(\nu) 2}\cdot\frac{2 p^{e^{\prime}}}{2} \}\\
     =&~ \sum_{S(\nu)\in\Hat{S}} \{ \Delta_{S(\nu) 1} + 2\Delta_{S(\nu) 2} \}\cdot\frac{ p^{e^{\prime}}}{2}\\
     =&~ \sum_{S(\nu)\in\Hat{S}} \frac{n^{\prime} p^{e^{\prime}}}{2}\quad\quad\quad\quad[\text{From Equation } (\ref{Equation32})]\\
     =&~ \frac{\mathcal{N} n}{2}
 \end{align*}
 So, the set $\{0,1,2,\ldots,\mathcal{N}\frac{n}{2}\}$ represents all the possible elements of the set $\mathcal{D}_{\frac{e}{2}}(n,2^e,\lambda)$, and therefore, $$\mid \mathcal{D}_{\frac{e}{2}}(n,2^e,\lambda)\mid~ = \frac{\mathcal{N} n}{2}+1.$$
 \end{enumerate}\qed
\end{proof}

The following result presents a sufficient condition for a constacyclic code over $\mathcal{A}$ to be Galois LCD.
\begin{theorem}
 Let $n$ be a positive integer relatively prime to $q$ and $\lambda = \sum_{S(\nu)\in\Hat{S}} \lambda_{S(\nu)}e_{S(\nu)}$ be a unit in $\mathcal A$ with $r = \mathrm{ord}(\lambda)$ is such that $\gcd(r,n) = 1$ and $\mathcal A$ divides $1+p^{e-k}$, where $0\leq k<e$. Then a $\lambda$-constacyclic code $C$ of length $n$ over $\mathcal A$ is a $k$-Galois LCD code if $nr\in B_1$.
\end{theorem}

\begin{proof}
    Let $\mathrm{ord}(\lambda_{S(\nu)}) = r_{S(\nu)}$ for $S(\nu)\in\Hat{S}$. Then from Lemma \ref{Lemma9}, $r_{S(\nu)}\mid r$ for each $S(\nu)\in\Hat{S}$. Also, $n$ being relatively prime to $q$, we have $n=n^{\prime}$ as per the notations that are used in the statement of Theorem \ref{Theorem13}.
    
    Let $nr\in B_1$. From the definition of the set $B_1$, it is easy to note that for each positive integer $j$ dividing $nr$, we get $j\in B_1$. This implies that $nr_{S(\nu)}\in B_1$ for each $S(\nu)\in\Hat{S}$. So, $\mathcal{A}_{S(\nu)}\subseteq B_1$ and $D_{S(\nu) t} = \mathcal{A}_{S(\nu)}\cap B_{a_t} = \emptyset$ for each $S(\nu)\in\Hat{S}$ and $2\leq t\leq s$. Then, from Corollary \ref{Corollary4}, $\mathrm{dim}_q(\mathrm{hull}_k(C)) = 0$. Thus, $C$ is a $k$-Galois LCD code.    
    \qed
\end{proof}

\section{Quantum Codes}\label{sec:quan}
 In this section, we derive entanglement-assisted quantum error-correcting codes (shortly, EAQECCs) as an application of our obtained results. Here, we use the Galois hull dimension to construct EAQECCs. Recall that by an $[[n,\mathrm{k},d;c]]_q$-EAQECC, we refer to a $q$-ary standard quantum code of length $n$, distance $d$, dimension $\mathrm{k}$, and the number of pre-shared entanglement $c$. Based on the hull dimension, EAQECCs can be constructed based on the next result.
\begin{lemma}\cite{Cao}
    For a linear code $C$ with parameters $[n,\mathrm{k},d]_q$ and $k$-Galois dual code $C^{\perp_k}$, there exists an EAQECC of parameters $[[n,\mathrm{k}-\dim(\mathrm{hull}_k(C)),d;n-\mathrm{k}-\dim(\mathrm{hull}_k(C))]]_q$.
\end{lemma}\label{lem.EAQC}

The following lemma provides a bound on the parameters of an EAQECC.
\begin{lemma}\cite{Lai}\label{Lemma8}
    For any $[[n,\mathrm{k},d;c]]_q$-EAQECC, if $d\leq \frac{n+2}{2}$ then we have
    \begin{align*}
        2(d-1) \leq n-\mathrm{k}+c.
    \end{align*}
\end{lemma}
If the parameters of an $[[n,\mathrm{k},d;c]]_q$-EAQECC achieve the bound given in Lemma \ref{Lemma8}, then we call it an \textit{EAQECC with relatively large minimum distance}.

\subsection{Gray Map}\label{Section4.1}
Let $GL_{\mathcal{N}}(\mathbb{F}_q)$ be the set of all invertible matrices of order $\mathcal{N}$ and transpose of $M\in GL_{\mathcal{N}}(\mathbb{F}_q)$ is denoted by $M^T$. Every $a\in \mathcal A$ can be expressed as $a = \sum_{S(\nu)\in \Hat{S}} a_{S(\nu)} e_{S(\nu)}$, where $a_{S(\nu)} \in \mathbb{F}_q$ for all $S(\nu)\in \Hat{S}$. Then, we define the map
\begin{equation}\label{eq:graymap}
\begin{array}{cccl}
    \psi:&~ \mathcal{A}&\longrightarrow &\mathbb{F}_q^{\mathcal{N}}\\
    &a&\mapsto&\psi(a)=(a_{S(\nu_1)},a_{S(\nu_2)},\dots,a_{S(\nu_\mathcal{N})})M,
    \end{array}
\end{equation}
where $\mid \Hat{S}\mid~ = \mathcal{N}$, and $M\in GL_{\mathcal{N}}(\mathbb{F}_q)$ is such that $M (M^{(p^k)})^T = \gamma I_{\mathcal{N}}$ with $\gamma \in \mathbb{F}_q^{\ast}$, where  $I_{\mathcal{N}}$ denotes the identity matrix of order $\mathcal{N}$, and if $M = (m_{ij}),$ then $M^{(p^k)} = (m_{ij}^{p^k})$, $0\leq k < e$. 

The  map in Equation~(\ref{eq:graymap}) can be extended componentwise over $\psi: \mathcal{A}^n\longrightarrow \mathbb{F}_q^{n\mathcal{N}}$. We define the Gray weight for an element $\boldsymbol{a}\in \mathcal{A}^n$ by $w_G(\boldsymbol{a})=w_H(\psi(\boldsymbol{a}))$ where $w_H$ refers the Hamming weight in $\mathbb{F}_q$. It is easy to verify that the map $\psi$ is bijective linear isometry from $\mathcal{A}^n$ (Gray distance) to $\mathbb{F}_q^{n\mathcal{N}}$ (Hamming distance).

\begin{lemma}\label{hull.supp}
    For a linear code $C$ over the ring $\mathcal{A}$, 
    \begin{enumerate}
        \item $\psi(C^{\perp_k})=\psi(C)^{\perp_k}$.
        \item $\psi(\mathrm{hull}_{k}(C)) = \mathrm{hull}_{k}(\psi(C))$.
    \end{enumerate}

\end{lemma}
\begin{proof}
\begin{enumerate}
\item From \cite{Fan}, we know that $C^{\perp_k} = (C^{(p^{e-k})})^{\perp}$, and from \cite{Islam1}, we have $\psi(C) = \psi(C)^{\perp}$. Here, $C^{(p^{e-k})} = \{c^{p^{e-k}} : c\in C\}$, and $(C^{(p^{e-k})})^{\perp}$ and $\psi(C)^{\perp}$ represent the Euclidean duals of $C^{(p^{e-k})}$ and $\psi(C)$, respectively. Then, we have
    \begin{align*}
        \psi(C^{\perp_k}) = \psi((C^{(p^{e-k})})^{\perp}) = \psi(C^{(p^{e-k})})^{\perp} = (\psi(C)^{(p^{e-k})})^{\perp} = \psi(C)^{\perp_k}.
    \end{align*}
\item First, we show that $\psi(C\cap  C^{\perp_k}) = \psi(C)\cap               \psi(C^{\perp_k})$. Take $\alpha \in \psi(C\cap C^{\perp_k})$. Then       there exists $c\in C\cap C^{\perp_k}$ such that $\alpha =  \psi(c)$. Now, $c\in C$ implies that $\psi(c) \in \psi(C)$, and $c\in C^{\perp_k}$ implies that $\psi(c) \in \psi(C^{\perp_k})$. Thus, $c\in C\cap C^{\perp_k}$           implies that $\alpha = \psi(c) \in  \psi(C)\cap \psi(C^{\perp_k})$, and hence $\psi(C\cap C^{\perp_k}) \subseteq \psi(C)\cap                    \psi(C^{\perp_k})$.

    Again, take $\beta \in \psi(C)\cap \psi(C^{\perp_k})$. Then there exist $c\in C$ and $c^{\prime} \in C^{\perp_k}$ such that $\psi(c) = \beta = \psi(c^{\prime})$. Since $\psi$ is an injective map, we have $c = c^{\prime}\in C\cap C^{\perp_k}$. Thus, $\beta = \psi(c) \in \psi(C\cap C^{\perp_k})$, and hence $\psi(C)\cap \psi(C^{\perp_k}) \subseteq \psi(C\cap C^{\perp_k})$. Therefore, $\psi(C\cap C^{\perp_k}) = \psi(C)\cap \psi(C^{\perp_k})$. Now, the result follows from part $1$.
    \end{enumerate}
\end{proof}
For a linear code $C$ of length $n$ over $\mathcal A$, let $\psi(C)=D$. Then $D$ is a linear code of length $n\mathcal{N}$. Again, from Lemma \ref{hull.supp}, we have $\dim(\mathrm{hull}_{k}(D))=\dim(\mathrm{hull}_{k}(\psi(C)))=\dim(\psi(\mathrm{hull}_{k}(C)))=\dim_q(\mathrm{hull}_{k}(C))$. Based on the hull dimension of $D$, we construct EAQECCs as follows:
\begin{theorem}\label{theorem.EAQE}
    For a linear code $C$ of length $n$ over $\mathcal A$, let $\psi(C)=D$ be an $[n\mathcal{N},\mathrm{k},d]$ linear code over $\mathbb{F}_q$. Then there exists a EAQECC of parameters $[[n\mathcal{N}, \mathrm{k} -\dim_q(\mathrm{hull}_{k}(C)),d; n\mathcal{N}-\mathrm{k}-\dim_q(\mathrm{hull}_{k}(C))]]_q$.
\end{theorem}

\begin{remark}\label{rem:sum} {Note that when $M=I_{\mathcal{N}}$  the identity matrix of order $\mathcal{N}$ in the Gray map in Equation~(\ref{eq:graymap}), the discussion in Section~\ref{sec.3} above shows that the resulting codes are the direct sum of
shorter codes related to the decomposition of the ring.
    Thus, the minimum distance is upper bounded by that of each
component, and for that parameter, there is no gain in using the rings. Anyway, the underlying algebraic structure for their definition allows a compact way for dealing with this type of codes as well as for their ring operations. Moreover, the ring structure adds some further symmetries (those provided by the defining ideal $I=\langle \{t_i(x_i)\}_{i=1}^\ell\rangle$) within the codes in field representation given by the Gray map other than those given by the constacyclic shift. These extra symmetries could help in decoding those larger codes. In other words, the resulting codes are not just direct sums of constacyclic codes but also $t_i(x_i)$-polycyclic codes ($i=1,\ldots, \ell$) when they are seen in $\mathbb F_q$.}
\end{remark}

\subsection{Examples}
In this subsection, we apply Theorem \ref{Theorem13} and the Gray map defined in Section \ref{Section4.1} to find the dimensions of the Galois hulls of constacyclic codes over $\mathcal A$. Then we use Lemma \ref{hull.supp} and Theorem \ref{theorem.EAQE} to obtain EAQECCs over $\mathbb{F}_q$.
\begin{example}\label{Example2}
Let us consider the ring $R = \mathbb{F}_9[x_1,x_2]/\langle t_1(x_1), t_2(x_2) \rangle$, where $t_1(x_1) = x_1^2 + x_1 + \omega$, $t_2(x_2) = x_2^2 + \omega^5 x_2 + 2$. Let $\mathbb{F}_{9} = \mathbb{F}_{3}(\omega)$ be the field with $9$ elements where $\omega^2 + 2\omega + 2 = 0$. Here, $p=3$, $e=2$, $\ell = 2$, $\mathcal{N} = 2$. Then, there are $2$ idempotents $e_{\alpha}$, $\alpha = 1,2$, in the orthogonal idempotent decomposition of $R$. Take $\lambda = e_1 + \omega^2 e_2$, $n=7$, $k=1$. Then $n^{\prime} = 7$, $l = 2$, and $r = 4$. Also, $\lambda_1 = 1$, $\lambda_2 = \omega^2$ and $r_1 = 1$, $r_2 = 4$. Let $a_1 = 1$, and $a_2 = 2$. Now, we have
    \begin{align*}
        \mathcal{A}_1 =&~ \Big\{j\in\mathbb{N} : \gcd(j,9) = 1,~ j \text{ divides } 7 \text{ and } \gcd\big(\frac{7}{j},1\big)=1\Big\} \\
        =&~ \{1, 7\};\\
        \mathcal{A}_2 =&~ \Big\{j\in\mathbb{N} : \gcd(j,9) = 1,~ j \text{ divides } 28 \text{ and } \gcd\big(\frac{28}{j},4\big)=1\Big\} \\
        =&~ \{4, 28\};\\
        B_1 =&~ \Big\{j\in\mathbb{N} : j\mid (9^{l_1}+3) \text{ for at least one non-negative integer } l_1\Big\}\\
        =&~ \{1,2,3,4,6,7, \ldots, 28, \ldots\};\\
        B_2 =&~ \Big\{j\in\mathbb{N}\setminus B_1 : j\mid (9^{l_1}-9) \text{ for at least one non-negative integer } l_1\Big\}\\
        =&~ \{5,8,9,10, \ldots\}.
    \end{align*}
The factorization of the polynomials $x^{7}-1$ and $x^{7}-\omega^2$ over $\mathbb{F}_{9}$ are respectively
\begin{alignat*}{9}
    & x^{7}-1 &&=&&~ (x+2) (x^3 + \omega x^2 + \omega^7 x + 2) (x^3 + \omega^3 x^2 + \omega^5 x + 2)\\
    \text{and}\quad\quad &x^{7}-\omega^2 &&=&&~ (x+\omega^2) (x^3 + \omega x^2 + \omega x + \omega^6) (x^3 + \omega^7 x^2 + \omega^3 x + \omega^6).
\end{alignat*}
Let $C$ be the $(e_1 + \omega^2 e_2)$-constacyclic code generated by $g(x) = \sum_{\alpha =1}^{2} e_{\alpha} g_{\alpha}(x)$, where $g_1(x) = x^3 + \omega^3 x^2 + \omega^5 x + 2$, $g_2(x) = x + \omega^2$. Therefore, $D_{11}=A_1\cap B_1=\{1,7\}$, $D_{21} = A_2\cap B_1=\{4,28\}$, $D_{12}=A_1\cap B_2=\emptyset = D_{22}$. Hence, $\text{ord}_1(9)=1, \text{ord}_7(9)=3, \text{ord}_4(9)=1, \text{ord}_{28}(9)=3$, $\beta_{11}(1)=1, \beta_{11}(7)=3, \beta_{21}(4)=1, \beta_{21}(28)=3$. Therefore, we have
\begin{alignat*}{9}
   & b_{111} && =&&~ b_{1111}\\
   & && =&&~ 1 - \max\{0,1-0\} = 0;\\
   & b_{117} && =&&~ b_{1117} + b_{1127}\\
   & && 
   =&&~ [1 - \max\{0,1-0\}] + [1 - \max\{1,1-1\}] = 0;\\
   & b_{214} && =&&~ b_{2114}\\
   & && 
   =&&~ 1 - \max\{1,1-1\} = 0;\\
   & b_{2128} && =&&~ b_{21128} + b_{21228}\\
   & && 
   =&&~ [1 - \max\{0,1-0\}] + [1 - \max\{0,1-0\}] = 0.
\end{alignat*}
Thus, the $9$-dimension of the $1$-Galois hull of $C$ is
\begin{align*}
    \dim_9(\text{hull}_1(C)) = (\text{ord}_1(9) \times b_{111}) + (\text{ord}_7(9) \times b_{117}) + (\text{ord}_4(9) \times b_{214}) + (\text{ord}_{28}(9) \times b_{2128}) = 0.
\end{align*}
So, by Lemma \ref{hull.supp}, we have $\dim_9(\text{hull}_1 (\psi(C)))=0$. Again, $\psi(C)$ has the parameters $[14, 10, 4]_9$ which is an optimal code \cite{Grassl}. For the Gray map $\psi$, we randomly choose an invertible matrix of order $2$, given by
    \begin{center}$
        M = \begin{bmatrix}
            \omega &~ \omega^3 \\
            \omega^7 & \omega^5
        \end{bmatrix}.$
    \end{center}
    Now, by Theorem \ref{theorem.EAQE}, we have an EAQECC of parameters $[[14, 10, 4; 4]]_9$.
\end{example}

\begin{example}\label{Example3}
Let us consider the ring $R = \mathbb{F}_{25}[x_1,x_2]/\langle t_1(x_1), t_2(x_2) \rangle$, where $t_1(x_1) = x_1^2 + x_1 + \omega^{13}$, $t_2(x_2) = x_2^2 + \omega^3 x_2 + 1$. Let $\mathbb{F}_{25} = \mathbb{F}_{5}(\omega)$ be the field with $25$ elements where $\omega^2 + 4\omega + 2 = 0$. Here, $p=5$, $e=2$, $\ell = 2$, $\mathcal{N} = 2$. So, there are $2$ idempotents $e_{\alpha}$, $\alpha = 1,2$, in the orthogonal idempotent decomposition of $R$. Take $\lambda = e_1 + \omega^8 e_2$, $n=13$, $k=1$. Then $n^{\prime} = 13$, $l = 2$, and $r = 3$. Also, $\lambda_1 = 1$, $\lambda_2 = \omega^8$ and $r_1 = 1$, $r_2 = 3$. Let $a_1 = 1$, and $a_2 = 2$. Now, we have
    \begin{align*}
        \mathcal{A}_1 =&~ \Big\{j\in\mathbb{N} : \gcd(j,25) = 1,~ j \text{ divides } 13 \text{ and } \gcd\big(\frac{13}{j},1\big)=1\Big\} \\
        =&~ \{1, 13\};\\
        \mathcal{A}_2 =&~ \Big\{j\in\mathbb{N} : \gcd(j,25) = 1,~ j \text{ divides } 39 \text{ and } \gcd\big(\frac{39}{j},3\big)=1\Big\} \\
        =&~ \{3, 39\};\\
        B_1 =&~ \Big\{j\in\mathbb{N} : j\mid (25^{l_1}+5) \text{ for at least one non-negative integer } l_1\Big\}\\
        =&~ \{1,2,3,5,6,7,9,10,14, \ldots\};\\
        B_2 =&~ \Big\{j\in\mathbb{N}\setminus B_1 : j\mid (25^{l_1}-25) \text{ for at least one non-negative integer } l_1\Big\}\\
        =&~ \{4,8,11,12,13,\ldots, 39, \ldots\}.
    \end{align*}
The factorization of the polynomials $x^{13}-1$ and $x^{13}-\omega^8$ over $\mathbb{F}_{25}$ are respectively
\begin{alignat*}{9}
    & x^{13}-1 &&=&&~ (x+4) (x^2 + \omega x + 1) (x^2 + \omega^2 x + 1) (x^2 + \omega^5 x + 1)\\
    & && &&~ \times(x^2 + \omega^{10} x + 1) (x^2 + \omega^{19} x + 1) (x^2 + \omega^{23} x + 1)\\
    \text{and}\quad\quad &x^{13}-\omega^8 &&=&&~ (x+\omega^20) (x^2 + \omega^3 x + \omega^{16}) (x^2 + \omega^7 x + \omega^{16}) (x^2 + \omega^9 x + \omega^{16})\\
    & && &&~ \times(x^2 + \omega^{10} x + \omega^{16}) (x^2 + \omega^{13} x + \omega^{16}) (x^2 + 3x + \omega^{16}).
\end{alignat*}
Let $C$ be the $(e_1 + \omega^8 e_2)$-constacyclic code generated by $g(x) = \sum_{\alpha =1}^{2} e_{\alpha} g_{\alpha}(x)$, where $g_1(x) = (x+4) (x^2 + \omega x + 1)$, $g_2(x) = x + \omega^{20}$. Thus, $D_{11}=A_1\cap B_1=\{1\}$, $D_{12}=A_1\cap B_2=\{13\}$, $D_{21} = A_2\cap B_1=\{3\}$, $D_{22}=A_2\cap B_2=\{39\}$. So, $\text{ord}_1(25)=1, \text{ord}_{13}(25)=2, \text{ord}_3(25)=1, \text{ord}_{39}(25)=2$, $\beta_{11}(1)=1, \beta_{12}(13)=3, \beta_{21}(3)=1, \beta_{22}(39)=3$. Hence, we have
\begin{alignat*}{9}
   & b_{111} && =&&~ b_{1111}\\
   & && =&&~ 1 - \max\{1,1-1\} = 0;\\
   & b_{1213} && =&&~ b_{12113} + b_{12213} + b_{12313}\\
   & && 
   =&&~ [2 - (\max\{1,1-0\} + \max\{0,1-1\})] + [2 - (\max\{0,1-0\} + \max\{0,1-0\})]\\
   & && &&~ + [2 - (\max\{0,1-0\} + \max\{0,1-0\})] = 1;\\
   & b_{213} && =&&~ b_{2113}\\
   & && 
   =&&~ 1 - \max\{1,1-1\} = 0;\\
   & b_{2239} && =&&~ b_{22139} + b_{22239} + b_{22339}\\
   & && 
   =&&~ [2 - (\max\{0,1-0\} + \max\{0,1-0\})] + [2 - (\max\{0,1-0\} + \max\{0,1-0\})]\\
   & && &&~ + [2 - (\max\{0,1-0\} + \max\{0,1-0\})] = 0.
\end{alignat*}
Thus, the $25$-dimension of the $1$-Galois hull of $C$ is
\begin{align*}
    \dim_{25}(\text{hull}_1(C)) =&~ (\text{ord}_1(25) \times b_{111}) + (\text{ord}_{13}(25) \times b_{1213})\\
    &~ + (\text{ord}_3(25) \times b_{213}) + (\text{ord}_{39}(25) \times b_{2239}) = 2.
\end{align*}
Hence, by Lemma \ref{hull.supp}, we have $\dim_{25}(\text{hull}_1 (\psi(C)))=2$. Again, $\psi(C)$ has the parameters $[26, 22, 4]_{25}$ which is a Near MDS code, i.e., $d = n - \mathrm{k}$. For the Gray map $\psi$, we randomly choose an invertible matrix of order $2$ over $\mathbb{F}_{25}$, given by
    \begin{center}$
        M = \begin{bmatrix}
            \omega &~ \omega^2 \\
            2 & \omega^{13}
        \end{bmatrix}.$
    \end{center} Finally, by Theorem \ref{theorem.EAQE}, we have an EAQECC of parameters $[[26, 20, 4; 2]]_{25}$.
\end{example} 

In Table \ref{Table1}, we obtain a few EAQECCs over $\mathcal{A}$. We obtain these codes from $\lambda$-constacyclic codes over $\mathcal{A}$ using the Gray map $\psi$. For $q=4$, we consider $\mathcal{A} = \mathbb{F}_4[x_1,x_2]/\langle x_1^2 + x_1 + \omega, x_2^2 + \omega x_2 + \omega\rangle$, where $\omega$ is a primitive element of $\mathbb{F}_4$, and randomly choose an invertible matrix of order $2$, given by
    \begin{center}$
        M = \begin{bmatrix}
            \omega &~ 0 \\
            0 & \omega^2
        \end{bmatrix}.$
    \end{center}    
    For $q=9$, we consider the same ring $\mathcal{A}$ and the same invertible matrix $M$ given in Example \ref{Example2}, and for $q=25$, we consider $\mathcal{A}$ and $M$ as given in Example \ref{Example3}.
    In the table, we use the identification of   polynomials 
    of degree at most $n-1$ with  $n$-tuples 
    as mentioned in Section \ref{Section2.2} for representing the polynomials $g_1(x)$ and $g_2(x)$. 
    
    An $[n,\mathrm{k},d]$-linear code is said to be a \emph{Maximum Distance Separable (MDS) code} if the relation $d = n-\mathrm{k}+1$ is satisfied, and we will say that it is  \emph{Near Maximum Distance Separable (Near MDS) code}  if $d = n-\mathrm{k}$. Furthermore, we call a linear code \emph{Optimal} if the code parameters match with those given in Grassl's code table \cite{Grassl}, and we use the term \emph{Near Optimal} if for the same values of $q$, $n$, and $\mathrm{k}$, the value of $d$ of our code is one less than that given in Grassl's code table \cite{Grassl}. 
    
    With the help of Theorem \ref{theorem.EAQE}, we obtain the parameters of the EAQECCs from the $1$-Galois hulls of codes. In Lemma \ref{Lemma8}, we have seen that $2(d-1) \leq n-\mathrm{k}+c$. The last column of Table~\ref{Table1} indicates how much the value $2(d-1)$ differs from $n-\mathrm{k}+c$. We use the notation $[[n,\mathrm{k},d;c]]_{q}^{\ast}$ when $(n-\mathrm{k}+c) - 2(d-1) = 0$, and $[[n,\mathrm{k},d;c]]_{q}^{\dagger}$ when $(n-\mathrm{k}+c) - 2(d-1) = 2$. As we mentioned after stating Lemma \ref{Lemma8}, each of the $[[n,\mathrm{k},d;c]]_{q}^{\ast}$-EAQECCs given in the table is an \textit{EAQECC with relatively large minimum distance}.

\begin{center}
	\begin{table}[ht]
		\caption{EAQECCs obtained from $\lambda$-constacyclic codes over $\mathcal{A}$}
		\renewcommand{\arraystretch}{1.5}
		\begin{center}\label{Table1}
			\begin{tabular}{|c|c|c|c|c|c|c|c|c|}
				\hline
			$n$ & $\lambda$ & $g_{1}(x)$ & $g_2(x)$ & $\psi(C)$ & Remark on $\psi(C)$ & EAQECC & Diff. \\
				\hline

            $5$ & $\omega e_1+e_2$ & 	$(\omega,1,1)$  & $(1,\omega,1)$ & $[10,6,3]_{4}$ & Near Optimal \cite{Grassl} & $[[10,2,3;0]]_{4}$ & $4$ \\

            $4$ & $\displaystyle\sum_{i=1}^2 e_i$ & 	$(\omega^2,\omega^7,1)$  & $(\omega^6,\omega,1)$ & $[8,4,4]_{9}$ & Near MDS & $[[8,4,4;4]]_{9}^{\dagger}$ & $2$ \\

            $5$ & $\displaystyle\sum_{i=1}^2\omega^2 e_i$ & 	$(2,\omega^5,1)$  & $(\omega^6,1)$ & $[10,7,4]_{9}$ & MDS & $[[10,5,4;1]]_{9}^{\ast}$ & $0$ \\


            $5$ & $\displaystyle\sum_{i=1}^2\omega^2 e_i$ & 	$(\omega^2,\omega^5,\omega^7,1)$  & $(2,\omega^3,1)$ & $[10,5,6]_{9}$ & MDS & $[[10,1,6;1]]_{9}^{\ast}$ & $0$ \\


			$5$ & $2e_1+e_2$ & 	$(1,2,1,2,1)$  & $(2,\omega^3,\omega^7,1)$ & $[10,3,8]_{9}$ & MDS & $[[10,1,8;5]]_{9}^{\ast}$ & $0$ \\


			$6$ & $\displaystyle\sum_{i=1}^2 e_i$ & 	$(2,2,0,1,1)$  & $(2,1)$ & $[12,7,4]_{9}$ & Near Optimal \cite{Grassl} & $[[12,5,4;3]]_{9}$ & $4$ \\

			$7$ & $e_1+\omega^2 e_2$ & 	$(2,\omega^5,\omega^3,1)$  & $(\omega^2,1)$ & $[14,10,4]_{9}$ & Optimal \cite{Grassl} & $[[14,10,4;4]]_{9}^{\dagger}$ & $2$ \\


            $7$ & $e_1+2e_2$ & 	$(1,\omega^7,2,\omega^5,1)$  & $(1,\omega^7,\omega^5,1)$ & $[14,7,6]_{9}$ & Near Optimal \cite{Grassl} & $[[14,7,6;7]]_{9}$ & $4$ \\

			$7$ & $e_1+2e_2$ & 	$(1,\omega^7,2,\omega^5,1)$  & $(1,2,1,2,1,2,1)$ & $[14,4,10]_{9}$ & Optimal \cite{Grassl} & $[[14,4,10;10]]_{9}^{\dagger}$ & $2$ \\

            $13$ & $e_1+\omega^8 e_2$ & $(4,\omega^5,\omega^{17},1)$  & $(\omega^{20},1)$ & $[26,22,4]_{25}$ & Near MDS & $[[26,20,4;2]]_{25}^{\dagger}$ & $2$ \\

            $16$ & $\omega^{16} e_1+e_2$ & $(\omega^5,\omega^{11},1)$  & $(\omega^{21},1)$ & $[32,29,3]_{25}$ & Near MDS & $[[32,27,3;1]]_{25}^{\dagger}$ & $2$ \\

            $20$ & $\omega^{8} e_1+e_2$ & $(\omega^8,\omega^{10},1)$  & $(1,1)$ & $[40,37,3]_{25}$ & Near MDS & $[[40,35,3;1]]_{25}^{\dagger}$ & $2$ \\

            $26$ & $\displaystyle\sum_{i=1}^2 e_i$ & 	$(2,1,1,0,1)$  & $(1,1)$ & $[52,47,3]_{9}$ & Near Optimal \cite{Grassl} & $[[52,44,3;2]]_{9}$ & $6$ \\

				\hline
			\end{tabular}
		\end{center}
	\end{table}
	\end{center}{}

\begin{remark}
    In Table \ref{Table1}, we see that there is an EAQECC with parameters $[[10,1,8;5]]_9$. For this EAQECC, even though $d > \frac{n+2}{2}$, its parameters achieve the bound given in Lemma \ref{Lemma8}. For this reason, we call it an \textit{EAQECC with relatively large minimum distance}.
\end{remark}

\section{Conclusion}
Hulls of linear codes are important in determining the complexity of the algorithms that are used for checking the permutation equivalence of two linear codes or finding the automorphism group of a linear code. It has also been found to be applicable to obtaining good entanglement-assisted quantum error-correcting codes. Due to these applications, researchers have extensively focused on studying the hulls and their properties. Hulls of different families of linear codes have yet to be investigated with respect to various types of finite rings. Serial ring being one of them. In this work, we have considered a semisimple ring with affine polynomial algebra representation  $\mathcal A = \mathbb{F}_q[X_1,X_2,\ldots,X_\ell]/ I$ as the code alphabet, where the ideal $I$ is generated by monic single-variable polynomials. The code alphabets used in some recent research \cite{Goyal,Tian,Yadav2}, can be obtained as particular cases of the ring $\mathcal A$. Moreover, the mentioned works involve either the Euclidean or the Hermitian inner product, whereas, in our work, we use a more general inner product, namely the Galois inner product. Here, we study the Galois duals and the Galois hulls and derive a formula for the dimensions of the Galois hulls of constacyclic codes over $\mathcal A$. Finally, as an application, we show how these codes can be used for constructing entanglement-assisted quantum error-correcting codes. As a future line of research,
the enumeration of non-isometric constacyclic codes for a given Galois hull dimension is an interesting problem. This problem is still open over $\mathcal A$. Therefore, it may be interesting to count the number of constacyclic codes for a given Galois hull dimension over $\mathcal A$.
	
	\section*{Acknowledgment}
	The first and fourth authors would like to thank the CSIR, Govt. of India (under grant no. 09/1023 (0030)/2019-EMR-I) and DST, Govt. of India (under SERB File Number: MTR/2022/001052, vide Diary No / Finance No SERB/F/8787/2022-2023 dated 29 December 2022), respectively for providing financial support. The third author acknowledges the financial support under Grant PID2022-138906NB-C21 funded by MICIU/AEI/10.13039/501100011033 and by ERDF/EU. 
	\section*{Declarations}
\noindent \textbf{Data Availability Statement}: The authors declare that [the/all other] data supporting the findings of this study are available within the article. Any clarification may be requested from the corresponding author, provided it is essential. \\
\noindent  \textbf{Competing interests}: The authors declare that there is no conflict of interest regarding the publication of this manuscript.\\
\textbf{Use of AI tools declaration}:
The authors declare that they have not used Artificial Intelligence (AI) tools in the creation of this article.

\end{document}